# Wave-based Neuromorphic Circuit Networks: Tunable 2D Transmission-Line Metamaterials

SHREY THAKKAR[†] AND ANTHONY GRBIC[†,*]

*Department of Electrical Engineering and Computer Science, University of Michigan, Ann Arbor, MI 48109, USA*

[†]*These authors contributed equally.*

*agrbic@umich.edu*

**Abstract:** Neuromorphic computing promises fast and energy-efficient information processing for emerging applications such as artificial intelligence. This paper presents neuromorphic processors based on wave-based programmable transmission-line (TLIN) metamaterials. Specifically, 2D reactive electrical networks are proposed, consisting of a grid of interconnected subwavelength TLIN-based unit cells (neurons) with tunable reactive elements. During inference, the input data is encoded using single-tone sources impressed onto the network, and circuit quantities are measured to decode the output prediction. Computation is performed through wave propagation and interference across the grid, with the learned input-output relationships stored in the tunable reactive elements. A key contribution of this work is a scalable training method based on *in-situ* backpropagation. The adjoint variable method is used to derive a physical (electrical) realization of the backpropagation algorithm that is typically used to compute the gradient of the objective loss function in digital neural networks. This formulation computes the gradient from voltage measurements of two steady-state excitations: the forward pass (inference) and the adjoint pass (error backpropagation). This enables efficient training since it is independent of the number of trainable parameters and avoids the simulation-reality gap. To demonstrate the effectiveness of this approach, wave-based neuromorphic circuit networks are trained for allostery and classification tasks, and the system's robustness to damage is shown. This work paves the way for self-learning systems based on wave-based neuromorphic analog circuit hardware.

## 1. Introduction

Machine learning has emerged as the dominant form of deployed artificial intelligence (AI) due to its ability to learn from large and high-dimensional datasets. In particular, deep learning techniques have contributed to its success [1]. Deep learning uses artificial neural networks (ANNs), which are inspired by the human brain. The brain achieves complex functionality through the sparse connectivity of billions of neurons and their weighted local synaptic interactions. Similarly, ANNs are composed of several interconnected layers, each with many degrees of freedom that govern their interactions. Training an ANN involves adjusting its internal parameters (weights) to minimize a scalar loss function so that it "learns" to capture complex patterns in the data. ANNs are predominantly digital and implemented on von Neumann computing architectures, which physically separate memory and processing. This introduces practical limitations on the energy efficiency and scalability of digital ANNs, which require enormous data throughput during training [2,3]. In contrast, neuromorphic computing [4], characterized by co-located processing and memory, has emerged as a promising paradigm for the efficient and scalable implementation of neural networks.

Many neuromorphic systems emulate neural dynamics with digital or mixed-signal circuits, such as spiking neural processors (inspired by synaptic interactions found in the brain) [4–6] and memristive crossbar accelerators (mimicking standard machine-learning algorithms) [7–9]. In contrast, physical neural networks process information directly through the intrinsic dynamics of physical systems, and have been implemented on optical and mechanical hardware platforms [10–14]. Such systems support distributed and parallelized processing and operate on analog signals. Recent wave-based optical neuromorphic computing systems have demonstrated that they can realize highly parallel, low-loss linear transformations over broad bandwidths with considerably higher energy efficiency than their digital counterparts [13].

Programmable wave-based electrical networks, referred to as wave-based neuromorphic circuit networks (WNCs), are proposed in this work. They offer a complementary and accessible platform for realizing physical neural networks, but are generally unexplored. WNCs can leverage mature fabrication technologies and can be readily integrated with CMOS-based analog control and measurement systems. This enables scalable and adaptive implementations of high-speed, low-loss physical neural networks within the existing electronic infrastructure. The incorporation of nonlinear and time-varying components, such as CMOS-compatible or back-end-of-line (BEOL) compatible ferroelectric varactors, further provides a pathway toward a nonlinear, highly expressive neuromorphic architecture. Prior implementations of neuromorphic circuits have mainly considered static electrical networks, including resistive networks [15] and memristive-based architecture [7,16]. While the latter approach demonstrates memory and nonlinearity, both are limited to scalar-valued operations. Conversely, wave-based operations are complex-valued since electromagnetic waves have both amplitude and phase. This inherently reduces the number of computations needed and allows for efficient encoding of input data using standard quadrature amplitude modulation (QAM) [17].

This work investigates wave-based neuromorphic circuit networks (WNCs) as a platform for neuromorphic computing. Specifically, 2D reactive, linear transmission-line (TLIN) metamaterials comprising a grid of interconnected subwavelength unit-cells (neurons) with tunable reactive elements are considered, as shown in Fig. 1. The metamaterials are excited by single-tone sources. The programmable elements are the tunable shunt capacitors within each unit cell, which serve as the trainable degrees of freedom (memory) that store the learned input-output associations. During inference, the input data is encoded using the single-tone sources impressed onto the network, and circuit quantities are used to decode the output prediction. Computation is performed through wave propagation and interference across the grid. WNCs promise higher energy efficiencies than static, resistive counterparts since inference relies on scattering from tunable susceptances rather than continuous currents through conductances. Like optical neural networks, they offer a highly parallel, fast, and scalable platform. Linear operations can be performed passively with minimal propagation loss by processing information through scattering. Although the medium is linear, complex-valued signals and their interference enable rich transformations between inputs and outputs [17]. Furthermore, nonlinear mappings can be implemented by using judiciously chosen physical quantities to encode inputs and measure outputs [18].

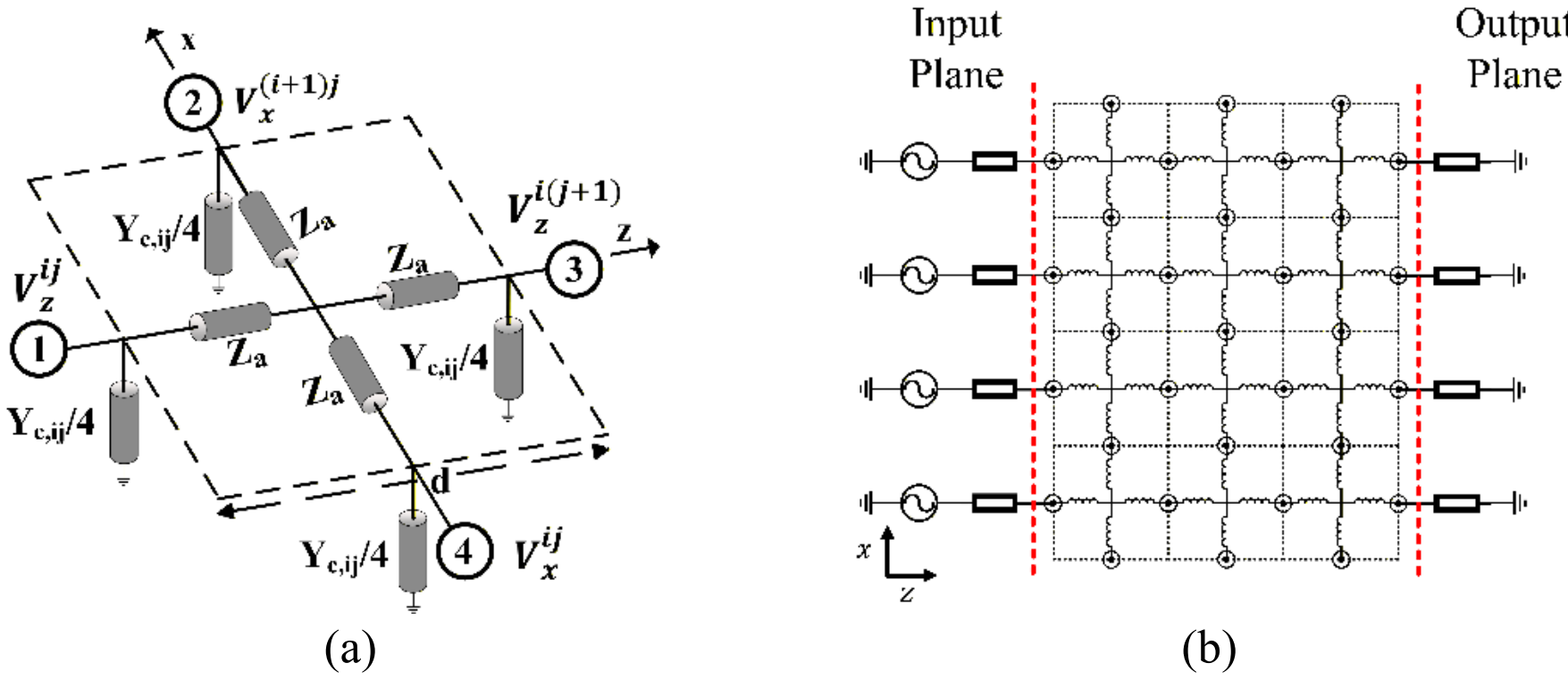

Fig. 1. Wave-based Neuromorphic Circuit Network (WNC). a) TLIN unit cell (neuron): four-port lossless transmission-line-based (TLIN) unit cell possessing fixed series inductors and tunable shunt capacitors. (b) 2D WNC consisting of an M×N (4×3) grid of TLIN unit cells. The input data is encoded as voltage source excitations at the input plane, and the network's prediction is decoded from the power through the terminations at the output plane.

This approach is demonstrated by training numerically simulated WNCs for machine-learning tasks. A fast 2D circuit network solver (CNS) [19,20] is used to rapidly simulate the WNC's response in the forward and adjoint passes. The WNC is represented as a large-scale admittance matrix of interconnected four-port admittance matrices representing each unit cell. The four-port admittance matrices are written in terms of the circuit elements comprising the unit cell (see Fig. 1(a)). Enforcing KCL at the interconnections between the unit cells yields a sparse linear system of equations that transforms impressed currents (data) into nodal voltages (prediction).

A key contribution of this work is a scalable training method based on *in-situ* backpropagation. Backpropagation is a standard algorithm for computing gradients of a loss function with respect to network parameters in digital neural networks. Here, the adjoint variable method is used to derive a physical (electrical) realization of the backpropagation algorithm. Rather than computing gradients through parameter perturbation, which scales linearly with the number of internal parameters, the proposed approach obtains the gradient of the objective loss function with respect to the trainable parameters in the WNC using only two steady-state excitations of the WNC. Specifically, the gradient is computed from nodal voltage measurements under forward excitation (inference) and backward/reverse excitation (error backpropagation) on the same WNC, using an error signal [8,10,11,14]. The adjoint-based in-situ backpropagation framework was first established for photonic neural networks in [14] and experimentally demonstrated in [10]. Here, we present the first derivation and adaptation of this framework to 2D spatially recurrent microwave circuit networks. The formulation of the gradient is a closed-form inner product of nodal voltages local to each tunable element. Although the formulation pertains to the proposed structure, it can be straightforwardly extended to non-isotropic unit cell topologies with multiple tunable elements as well.

This *in-situ* training is crucial for hardware realizations, as it enables scalable training independent of the number of trainable parameters and avoids the simulation-reality gap that arises when digital (*in-silico*) backpropagation is performed. The effectiveness of the proposed WNCs is illustrated through representative tasks such as allostery and classification. The retrainability of the WNC under damage and task-switching is also demonstrated. In this work, the *in-situ* training algorithm is validated in simulation using the CNS, and the path to hardware implementation is discussed in Section 5.

The article is organized as follows. Section 2 describes the TLIN unit cell and the 2D frequency-domain CNS used to simulate the WNCs (2D TLIN metamaterials) in both forward and *in-situ* backpropagation passes. Section 3 discusses the formulation of the training goal and *in-situ* computation of the gradient of the objective loss function using the adjoint variable method. Section 4 presents illustrative examples, including allostery and classification tasks. Concluding remarks follow.

## 2. 2D Circuit Network Solver

The constituent neurons of the physical neural network are the linear, lossless (TLIN) unit cells of the WNC, possessing series impedances $Z_a$ and shunt admittances $Y_{\mathrm{C},ij}/4$, as depicted in Fig. 1 (a), where *i* and *j* refer to the row and column address of the unit cell. The shunt element in each unit cell is a variable capacitor that represents each neuron's programmable memory and a trainable degree of freedom. On the other hand, the series impedances of all unit cells are fixed inductances, which can simply be the distributed inductance due to electrically-short transmission-line interconnects. Thus, $Y_{\mathrm{C},ij} = j\omega \mathrm{C}_{ij}$ and $Z_a = j\omega L$, where $\omega$ is the radian frequency at 10 GHz, $\mathrm{C}_{ij}$ is the tunable shunt capacitance, and $L$ is the series inductance set to 0.75 nH for all unit cells, corresponding to the inductance of subwavelength, metallic interconnects. The WNC comprises several of these unit cells (neurons) interconnected at their ports to form a rectangular lattice, as depicted in Fig. 1(b). The phase delay ($\phi$) across each isotropic unit cell can be written as

$$\sin^2\left(\frac{\phi_{ij}}{2}\right) = \frac{\omega^2 L C_{ij}}{2}. \tag{1}$$

When the TLIN unit cells are sufficiently subwavelength ($\lambda_0 \gg d$, where $\lambda_0$ is the free-space wavelength and $d$ the unit cell's dimension), they can be characterized in terms of a series impedance tensor and scalar admittance

$$\bar{\bar{Z}} = \begin{bmatrix} 2Z_a & 0 \\ 0 & 2Z_a \end{bmatrix}, \ Y_{C,ij}. \tag{2}$$

A relationship between an isotropic, homogeneous medium and a periodic network composed of the TLIN unit cell in Fig. 1(a) can be readily established as [21]

$$j\omega d \begin{bmatrix} \mu & 0 \\ 0 & \mu \end{bmatrix} = \begin{bmatrix} 2Z_a & 0 \\ 0 & 2Z_a \end{bmatrix}, \tag{3a}$$

$$j\omega d \epsilon_{ij} = Y_{C,ij}. \tag{3b}$$

Thus, Eq. (1) and Eq. (3b) indicate that tuning the capacitor is equivalent to tuning the phase delay experienced across the unit cell, or equivalently, varying the unit cell's effective permittivity.

A frequency-domain 2D circuit network solver (CNS) [19,20] is used to simulate the WNC's response (voltages and currents at the ports of the unit cells) during inference (the forward solve) and backpropagation of the error signal. The CNS models WNCs (or, more generally, 2D metamaterials) as a large-scale admittance matrix that relates impressed currents (input data) to nodal voltages (predictions). The large-scale admittance matrix is constructed by interconnecting four-port admittance matrices, one for each TLIN unit cell, and enforcing KCL at each node between unit cell ports. The four-port admittance-matrix entries of the TLIN unit cell shown in Fig. 1(a) can be written in terms of its circuit elements

$$Y_{11} = Y_{22} = Y_{33} = Y_{44} = \frac{1}{4}\left(Y_{C,ij} + \frac{3}{Z_a}\right), \tag{4a}$$

$$Y_{21} = Y_{31} = Y_{41} = Y_{32} = Y_{42} = Y_{43} = -\frac{1}{4Z_a}, \tag{4b}$$

where the numeric subscripts refer to the unit cell ports (nodes), while the alpha subscripts refer to the unit cell row and column address in the overall grid of unit cells. A WNC consisting of an M×N grid of TLIN unit cells is shown in Fig. 1(b). Boundary conditions are imposed using lumped impedances, and excitations are impressed as voltage sources with a source resistance. These sources are represented by their Norton equivalents, resulting in current sources (with a shunt source conductance) impressed at the corresponding nodes. The shunt source admittance is absorbed into the large-scale admittance matrix.

As noted, Kirchhoff's current law is enforced at the interconnections between four-port admittance-matrices representing the unit cells, yielding a sparse linear system of $n = 2\mathrm{MN} + \mathrm{M} + \mathrm{N}$ equations relating nodal voltages to impressed Norton current sources as [19],

$$\bar{\bar{Q}}\,\mathbf{v} = \mathbf{s}\,, \tag{5}$$

where $\bar{\bar{Q}} \in \mathbb{C}^{n\times n}$ is the large-scale admittance (interaction) matrix that captures the internal structure of the WNC and the unit-cell interactions, $\mathbf{v}$ is a $\mathbb{C}^{n\times 1}$ vector containing all the nodal voltages (at the ports of the unit cells) in the network, and $\mathbf{s}$ is a $\mathbb{C}^{n\times 1}$ vector of Norton current sources impressed on the WNC. Solving Eq. (5) yields nodal voltages (at ports of the unit cells) throughout the network. Further details on the construction of the system of equations in Eq. (5) as well as the form of $\bar{\bar{Q}}$ in terms of each unit cell's four-port admittance matrix (Eq. (4)) are provided in Supplement 1.

The CNS can be used to analyze any device composed of unit cells that support a single-mode. For instance, the CNS has been previously employed to analyze spatially-varying anisotropic media under TEM excitation [20] and planar microstrip networks composed of realizable TLIN unit cells [19]. The CNS can be extended to unit-cell geometries that exhibit higher-order intercell interactions (arbitrarily shaped unit-cell inclusions, such as metallic pillars [22,23]) by incorporating multiport network models (multimodal admittance matrices) and employing multimodal network theory [24]. This extension enables accurate *in-silico* training and evaluation of neuromorphic devices with complex internal structure prior to fabrication. For instance, all-metallic metamaterials composed of pillars within a parallel-plate waveguide [23] could realize artificial neural networks (ANNs) that are particularly suitable for space and high-power applications. Such ANNs promise low loss, high resilience to environmental conditions, and high power-handling capability. Furthermore, multi-harmonic models can be incorporated into the CNS in a similar manner, enabling in-silico training and evaluation of nonlinear WNCs, as well as those exploiting time-varying elements for frequency multiplexing to increase speed and parallelism and to overcome connectivity issues.

## 3. *In-Situ* Training Based on the Adjoint Variable Method

### 3.1 Framing the Learning Goal

The WNC can be characterized as a linear operator that transforms inputs (encoded using single-tone voltage sources) into outputs (circuit quantities), as described by Eq. (5). This transformation is referred to as inference, or the "forward solve". Training the WNC involves repeatedly performing inference on training data and minimizing an objective loss function by progressively updating the trainable parameters. The loss function $g$ assigns a scalar figure of merit to the device's performance as a function of the output nodal voltages that are, in turn, a function of the trainable parameters. We can concisely formulate the task learning goal as

$$\begin{gathered}\arg\min_{\mathbf{C}} g(\mathbf{C}) \\ \text{s.t. } \bar{\bar{Q}}\,\mathbf{v} = \mathbf{s} \\ \text{and } \mathbf{C}_{\text{lb}} \le \mathbf{C} \le \mathbf{C}_{\text{ub}}\end{gathered} \tag{6}$$

where $\mathbf{C}$ is a $\mathbb{R}^{(dof\cdot \mathrm{M}\cdot \mathrm{N})\times 1}$ vector of the tunable capacitances (trainable parameters) of the network, $dof = 1$ is the number of trainable degrees of freedom available per unit cell, $g(\mathbf{C})$ is an arbitrary loss function that is differentiable with respect to the nodal voltages, and $\mathbf{C}_{\text{lb}}$ and $\mathbf{C}_{\text{ub}}$ are vectors containing the lower and upper bounds of the available range of capacitance values. In other words, Eq. (6) states that the network's training entails determining the set of network capacitances that minimizes $g(\mathbf{C})$ subject to specified constraints. The training data is supplied to the network as $K$ input-output pairs, and a loss $g_k(\mathbf{C})$ is evaluated for the $k$th pair. The performance of the device over the training data is determined by summing over $k$ as

$$g(\mathbf{C}) = \sum_{k=1}^{K} g_k(\mathbf{C}). \tag{7}$$

The multi-categorical cross-entropy-based loss function

$$g_k(\mathbf{C}) = \frac{1}{2}\log\left(\frac{V_{label}^k(\mathbf{C}) \cdot V_{label}^{k}{}^*(\mathbf{C})}{\sum_o V_o^k(\mathbf{C}) \cdot V_o^{k*}(\mathbf{C})}\right) \tag{8}$$

is imposed for learning tasks. In Eq. (8), $V_{label}^k$ refers to the complex nodal voltage at the output node assigned to the *k*th training label, and $\sum_o V_o^k(\mathbf{C}) \cdot V_o^{k*}(\mathbf{C})$ is the sum of output power over all output nodes.

The WNC described here realizes a passive, reciprocal, linear complex-valued transformation between its input and output ports, with the square-law power readout as the only source of nonlinearity. This architecture is naturally suited to tasks that are linearly separable in the complex-valued feature space, which is strictly larger than real-valued linear separability. Tasks requiring deeper nonlinearity can be addressed through different strategies involving encoding nonlinearity [18,25], nonlinear elements, and higher-order inter-element coupling, which introduces structural nonlinearity [26]. Recent work by the authors has shown that higher-order coupling between unit cells can be captured by incorporating multimodal network theory into the CNS [27].

### *3.2* In-situ *Computation of the Gradient Based on the Adjoint Variable Method*

Training the network entails progressively updating $\mathbf{C}$ according to the gradient of the objective loss function with respect to the training parameters, $\nabla_{\mathbf{C}} g$. Evaluating the gradient by perturbing each capacitor and re-evaluating the loss function is a computationally expensive approach that scales linearly with the number of trainable parameters. Instead, the proposed in-situ training of the WNC relies on backpropagation of the prediction error via the physical process of wave propagation on the same network, combined with local rules. Backpropagation amounts to solving an adjoint problem, which conveniently reduces the gradient computation to a single additional steady-state solution, regardless of the number of training parameters. The manner in which the local rule combines the nodal voltage solutions from the forward pass and backpropagation to derive the gradient with respect to each capacitor $\mathbf{C}_{ij}$ is shown in Fig. 2 and will be described shortly.

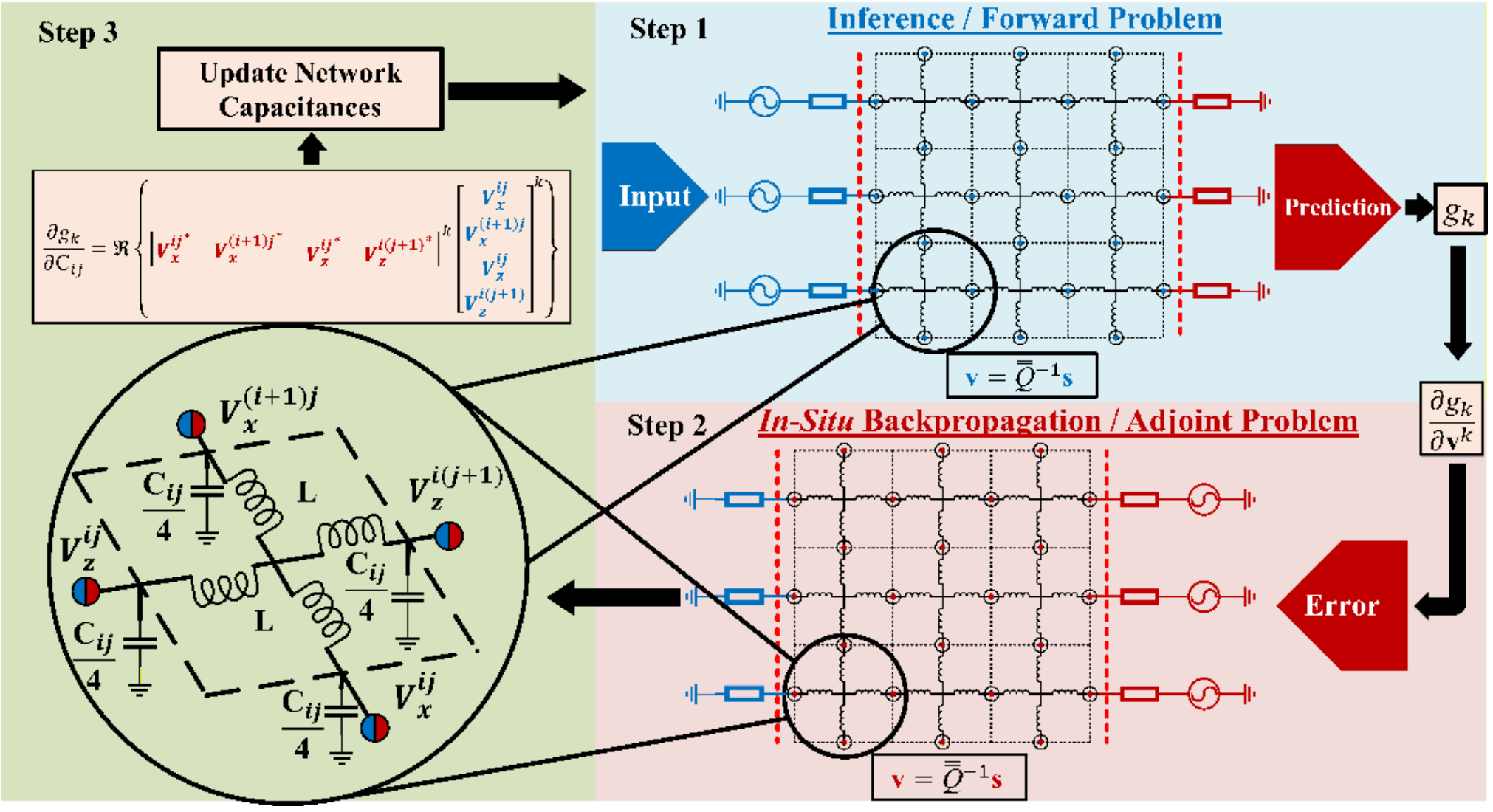

Fig. 2. Illustration of Wave-based Neuromorphic Circuit Network (WNC) training through *in-situ* backpropagation. The input data is encoded using voltage sources connected to the grid, and the computation is performed via wave propagation and scattering from tunable shunt capacitances in each unit cell. The output power is measured to decode the WNC's prediction, compute the loss function, and obtain the resulting error signal. In-situ backpropagation is performed physically by exciting the same network in reverse with the error signal (adjoint problem). The gradient of the loss function with respect to the tunable capacitances is computed as the inner product of the local nodal voltages measured after the forward pass and the *in-situ* backpropagation.

First, the gradient $\nabla_{\mathrm{C}} g$ is expressed as

$$\nabla_{\mathrm{C}} g = \sum_{k=1}^{K} \nabla_{\mathrm{C}} g_k = \sum_{k=1}^{K} \frac{\partial g_k}{\partial \mathbf{C}} = \sum_{k=1}^{K} \frac{\partial g_k}{\partial \mathbf{v}^k} \cdot \frac{\partial \mathbf{v}^k}{\partial \mathbf{C}}. \tag{9}$$

Since the loss $g_k$ is scalar and a function of the output nodal voltages (as seen from Eq. (8)), $\partial g_k / \partial \mathbf{v}^k$ is a Jacobian matrix that can be analytically calculated, with non-zero entries corresponding to the output nodes. The second term is a 2D Jacobian $\partial \mathbf{v}^k / \partial \mathbf{C} \in \mathbb{C}^{n \times (\mathrm{M} \times \mathrm{N})}$ (gradient of all nodal voltages with respect to tunable capacitances) that can be written using Eq. (5) as

$$\frac{\partial \mathbf{v}^k}{\partial \mathbf{C}} = -\bar{\bar{Q}}_k^{-1} \frac{\partial \bar{\bar{Q}}_k}{\partial \mathbf{C}} \mathbf{v}^k = -\bar{\bar{Q}}_k^{-1} \left( \frac{\partial \bar{\bar{Q}}_k}{\partial \mathrm{C}_{11}} \mathbf{v}^k \middle| \frac{\partial \bar{\bar{Q}}_k}{\partial \mathrm{C}_{12}} \mathbf{v}^k \middle| \cdots \middle| \frac{\partial \bar{\bar{Q}}_k}{\partial \mathrm{C}_{\mathrm{MN}}} \mathbf{v}^k \right). \tag{10}$$

The gradient of all nodal voltages with respect to one arbitrary tunable capacitance can be written as $\partial \mathbf{v}^k / \partial \mathrm{C}_{ij} \in \mathbb{C}^{n \times 1}$

$$\frac{\partial \mathbf{v}^k}{\partial \mathrm{C}_{ij}} = -\bar{\bar{Q}}_k^{-1} \left( \frac{\partial \bar{\bar{Q}}_k}{\partial \mathrm{C}_{ij}} \mathbf{v}^k \right). \tag{11}$$

The term $\partial\overline{\overline{Q}}_k/\partial \mathrm{C}_{ij} \in \mathbb{C}^{n\times n}$ is the sensitivity of the large-scale admittance matrix to a tunable capacitance and can be solved analytically if analytic expressions for the admittance matrix elements, such as Eq. (4), are available. Alternatively, the admittance matrix elements can be represented as continuously differentiable functions of the training parameters, so that Eq. (10) can be approximated using finite differences as in [19]. Computing $\partial\mathbf{v}^k/\partial \mathrm{C}_{ij}$ for each capacitor is typically performed *in-silico* and can be computationally expensive. However, in this work, it is greatly simplified and can be implemented in hardware (*in-situ*) using measured nodal voltages based on two key observations. First, the sensitivity of the four-port admittance matrix entries of a unit cell $Y^{ij}$ with respect to its capacitance $\mathrm{C}_{ij}$ can be derived using Eq. (4) as

$$\frac{\partial Y_{pq}^{ij}}{\partial \mathrm{C}_{ij}} = \begin{cases} \frac{j\omega}{4}, & p = q \\ 0, & p \neq q \end{cases}, \tag{12}$$

where $p, q \in \{1, 2, 3, 4\}$ refer to the unit cell ports (Fig. 1(a)). Second, it is evident from Eq. (S2)-(S7) in the supplementary that the self-terms exist only on the diagonal of the large-scale admittance matrix $\overline{\overline{Q}}_k$ in the rows corresponding to the KCL equations applied at the nodes ($V_x^{ij}$, $V_x^{(i+1)j}$, $V_z^{ij}$, and $V_z^{i(j+1)}$) of the unit cell with location (*i*, *j*). Therefore, $\partial\overline{\overline{Q}}_k/\partial \mathrm{C}_{ij}$ is a diagonal matrix having four non-zero entries in these rows, and we can simplify the bracketed term in Eq. (11) as

$$\frac{\partial\overline{\overline{Q}}_k}{\partial \mathrm{C}_{ij}}\mathbf{v}^k = \frac{j\omega}{4}\begin{bmatrix} 0 & & & \cdots & & & 0 \\ & \ddots & & & & & \\ & & 1 & & & & \\ & & & 1 & & & \\ \vdots & & & & \ddots & & \vdots \\ & & & & & 1 & \\ & & & & & \ddots & \\ & & & & & & 1 \\ & & & & & & \ddots \\ 0 & & & \cdots & & & 0 \end{bmatrix}\begin{bmatrix} V_x^{11} \\ \vdots \\ V_x^{ij} \\ V_x^{(i+1)j} \\ \vdots \\ V_z^{ij} \\ \vdots \\ V_z^{i(j+1)} \\ \vdots \\ V_z^{\mathrm{M(N+1)}} \end{bmatrix}^k. \tag{13}$$

Now, from Eq. (9) and Eq. (10),

$$\nabla_{\mathbf{C}} g_k = -\frac{\partial g_k}{\partial \mathbf{v}^k}\overline{\overline{Q}}_k^{-1}\frac{\partial\overline{\overline{Q}}_k}{\partial \mathbf{C}}\mathbf{v}^k, \tag{14}$$

which can be written as

$$\nabla_{\mathbf{C}} g_k = \mathbf{v}_{bp}^{k\ H}\frac{\partial\overline{\overline{Q}}_k}{\partial \mathbf{C}}\mathbf{v}^k, \tag{15}$$

where $\boldsymbol{v}_{bp}^k$ is expressed as the solution to the adjoint problem

$$\overline{\overline{Q}}_k^H\mathbf{v}_{bp}^k = -\left(\frac{\partial g_k}{\partial \mathbf{v}^k}\right)^H, \tag{16}$$

and $H$ indicates the conjugate transpose. The impressed current sources $-\left(\partial g_k/\partial \mathbf{v}^k\right)^H$ in the adjoint problem are a function of the error and are non-zero only at the output nodes, as previously noted. Physical variables like voltages and the unitary operator $\overline{\overline{Q}}_k$ are complex-conjugated in Eq. (16), indicating a time reversal. Therefore, $\mathbf{v}_{bp}^k$ is a vector of the physical nodal voltages of the same network under a backpropagation (time reversal) of the error. Now, $\partial g_k/\partial \mathrm{C}_{ij}$ (the gradient of the loss with respect to a capacitance for the $k$th training sample) can be expressed using Eq. (11) and Eq. (15) as

$$\frac{\partial g_k}{\partial \mathrm{C}_{ij}} = \mathbf{v}_{bp}^{k\ H} \frac{\partial \overline{\overline{Q}}_k}{\partial \mathrm{C}_{ij}} \mathbf{v}^k, \tag{17}$$

Finally, the gradient information for each capacitor $\mathrm{C}_{ij}$ over all of $K$ training samples can be expressed by substituting Eq. (13) into Eq. (17)

$$\frac{\partial g}{\partial \mathrm{C}_{ij}} = 2 \cdot \Re \left\{ \frac{j\omega}{4} \sum_{k=1}^{K} \left[ V_{bp,x}^{ij\ *} \quad V_{bp,x}^{(i+1)j^*} \quad V_{bp,z}^{ij\ *} \quad V_{bp,z}^{i(j+1)^*} \right]^k \begin{bmatrix} V_{fp,x}^{ij} \\ V_{fp,x}^{(i+1)j} \\ V_{fp,z}^{ij} \\ V_{fp,z}^{i(j+1)} \end{bmatrix}^k \right\}. \tag{18}$$

Thus, the gradient information for each capacitor is determined by evaluating the inner product of the associated unit cell's nodal voltages from the forward pass $\left[V_{fp,x}^{ij} \quad V_{fp,x}^{(i+1)j} \quad V_{fp,z}^{ij} \quad V_{fp,z}^{i(j+1)}\right]$ with the backpropagated/adjoint pass $\left[V_{bp,x}^{ij\ *} \quad V_{bp,x}^{(i+1)j^*} \quad V_{bp,z}^{ij\ *} \quad V_{bp,z}^{i(j+1)^*}\right]$. This is analogous to recent demonstrations of *in-situ* backpropagation in mechanical neural networks [11]. Since the unit cells are restricted to having lossless capacitors, only the real part is considered in (18). The resulting gradient agrees exactly with the finite-difference gradient in the small limit. Thus, the exact gradient information, regardless of the number of parameters, is obtained from just two independent steady-state solutions, as shown in Fig. 2. This is unlike well-established learning methods derived from energy-based models, such as equilibrium propagation [28] and coupled learning [29], that estimate the gradient using two dependent equilibrium states. The gradient is supplied to the constrained optimization routine fmincon() in MATLAB, which updates the capacitances. This result can be compared to the in-situ gradient formulation of [14], where the gradient is given by the overlap of the forward and adjoint fields over tunable phase-shifters in a bi-directional photonic network. Furthermore, the in-situ gradient formulation in [14] requires an additional forward solve to interfere the two fields and local intensity measurements for each tunable element, which must be subtracted to recover the gradient.

To extend this approach to arbitrary unit cells that have variable elements within it, a diagonalization of $\partial\overline{\overline{Q}}_k/\partial \boldsymbol{p}$ would be required, where $\boldsymbol{p}$ refers to the set of arbitrary tunable elements (training parameters), using an incidence matrix. Alternatively, a port can be defined across each variable element in the unit cell, which again would place non-zero elements of $\partial\overline{\overline{Q}}_k/\partial \boldsymbol{p}$ only along its diagonal. This extension would allow up to 10 degrees of freedom to be incorporated per unit cell (entries in the upper right triangle of a lossless, reciprocal four-port admittance matrix), enabling the design of WNCs with considerably smaller footprints.

## 4. Results

### *4.1 Allostery in Wave-based Neuromorphic Circuits*

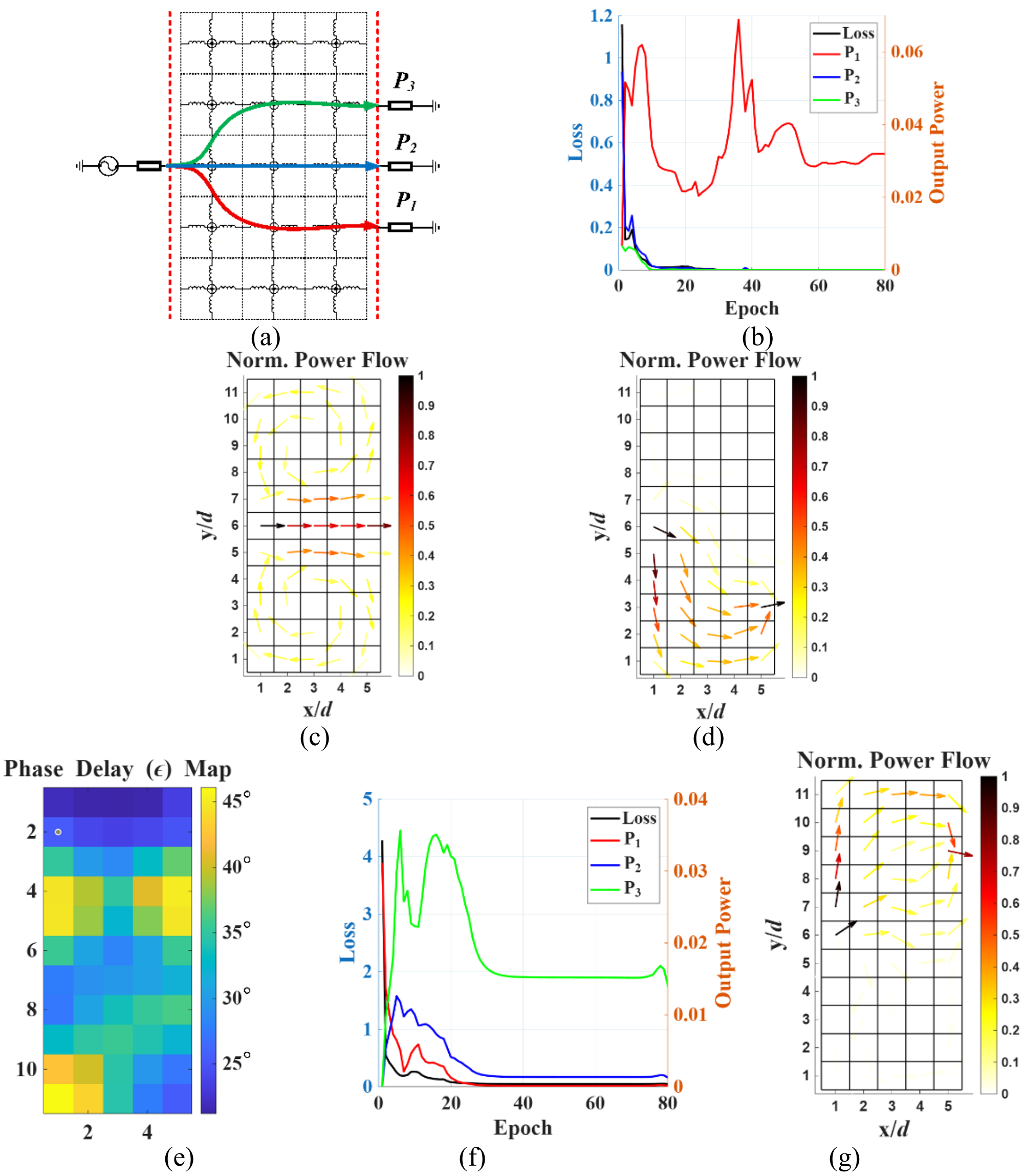

Fig. 3. a) Allostery in a Wave-based Neuromorphic Circuit Network (WNC) through directed power flow. b) Evolution of loss function and power through output nodes with epoch. c) Normalized power flow in the initial homogeneous WNC. d) Normalized power flow in the trained WNC, which directs power through output node $P_1$. e) Phase delay distribution of the trained WNC. f) Evolution of loss function and power through output nodes with epoch of the retrained WNC. g) Normalized power flow in the retrained WNC which directs power through output node $P_3$.

Directed transport or routing of signals between spatially distinct regions through local internal couplings is a canonical task in neuromorphic systems. This behavior is closely related to the phenomenon of allostery observed in biological systems, which refers to the ability of a structured system to transmit information between spatially distinct sites through its internal couplings [30,31]. For example, in proteins [32], the binding of a molecule at a binding site (input signal) causes a conformational change elsewhere (active site) that regulates the protein's activity. Allostery in biological systems is also related to 'transport control', which encompasses phenomena such as neural signal routing and blood flow redistribution. Neural signal routing refers to the process in neural systems through which the excitation of a particular neuron leads to the preferential excitation of one downstream neuron population over another due to synaptic weights, the system's connectivity, and the present state.

The next example demonstrates allostery in WNCs (trained through *in-situ* backpropagation), whereby power from a symmetrically placed voltage excitation at the input plane is routed through one of three output nodes, as illustrated in Fig. 3(a). A 10 V voltage source is connected to an 11×5 WNC through a source resistance of 175 Ω, and the three output nodes are terminated with 175 Ω as well. The remaining peripheral/boundary nodes are left open-circuited. The WNC maps input voltage excitations to output power flow, thereby introducing nonlinearity into the input-output relationship. First, the WNC is trained using the cross-entropy loss function given by Eq. (8), to direct power to the output node indicated by the red arrow. Fig. 3(b) plots the loss function and output power through each node ($P_1$ (red), $P_2$ (blue), and $P_3$ (green)) at each epoch. Initially, the loss function decreases rapidly, and then settles around epoch 20. Since the grid was initialized as isotropic and homogeneous, $P_2$ is the highest and $P_1$ is identical to $P_3$ at epoch 0. The normalized power flow within the isotropic, homogeneous WNC is plotted in Fig. 3(c). As the network is trained, power is disproportionately directed to $P_1$, while the powers to $P_2$ and $P_3$ are minimized (Fig. 3(d)). Note that at epoch 36, power through $P_1$= 0.7 W, which is the maximum available power from the source, indicating that the WNC can learn to impedance-match if the loss function incorporates a penalty for impedance mismatch. Since tuning a capacitance is equivalent to tuning the phase delay across a unit cell (as indicated by Eq. (1)) or tuning the permittivity (as indicated by Eq. (3b)), the optimized WNC is plotted in Fig. 3(e) as a function of the phase delay across each unit cell.

It is imperative that an ANN can be retrained to achieve a different task. Here, the WNC is retrained to redirect the input power to $P_3$ while minimizing it to $P_1$ and $P_2$, using the optimized distribution in Fig. 3(e) as a seed. Fig. 3(f) shows that initially the power through the output nodes matches that at Epoch 80 of Fig. 3(b), but then rapidly evolves as the loss is minimized. The power flow within the retrained WNC is shown in Fig. 3(g).

### *4.2 Classification on the Palmer Penguin Dataset with Amplitude and Phase Encoding*

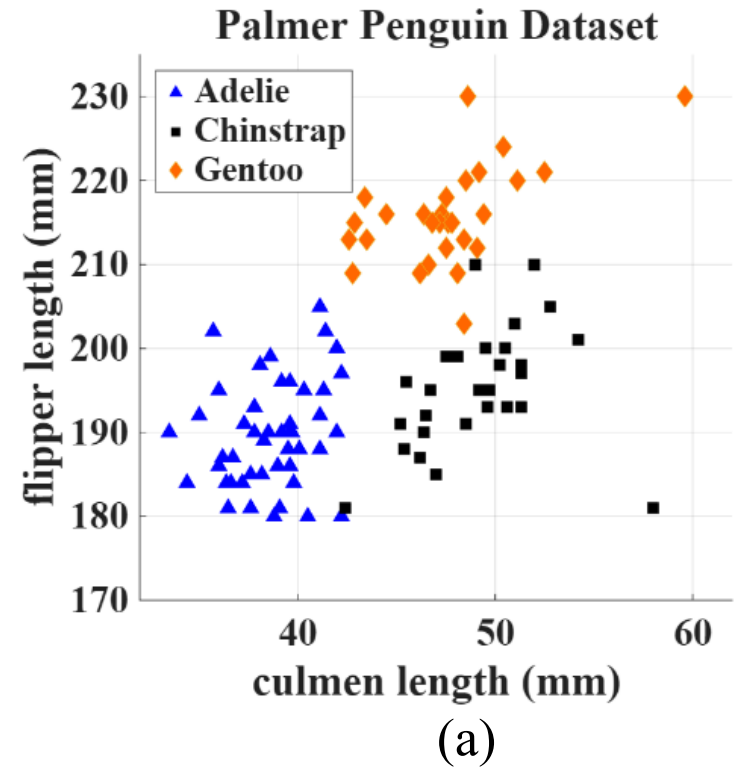

(a)

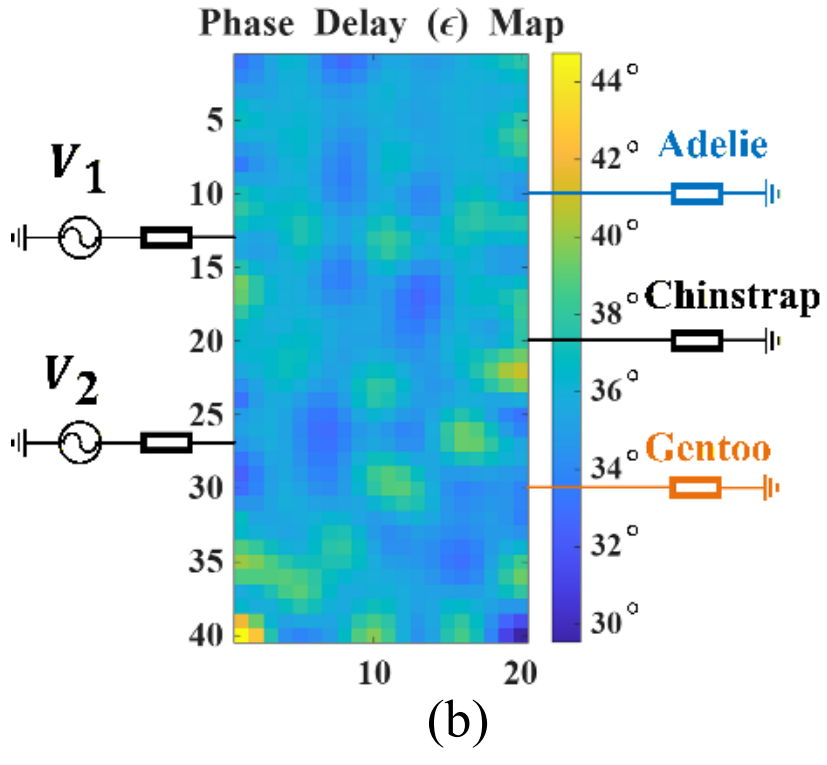

(b)

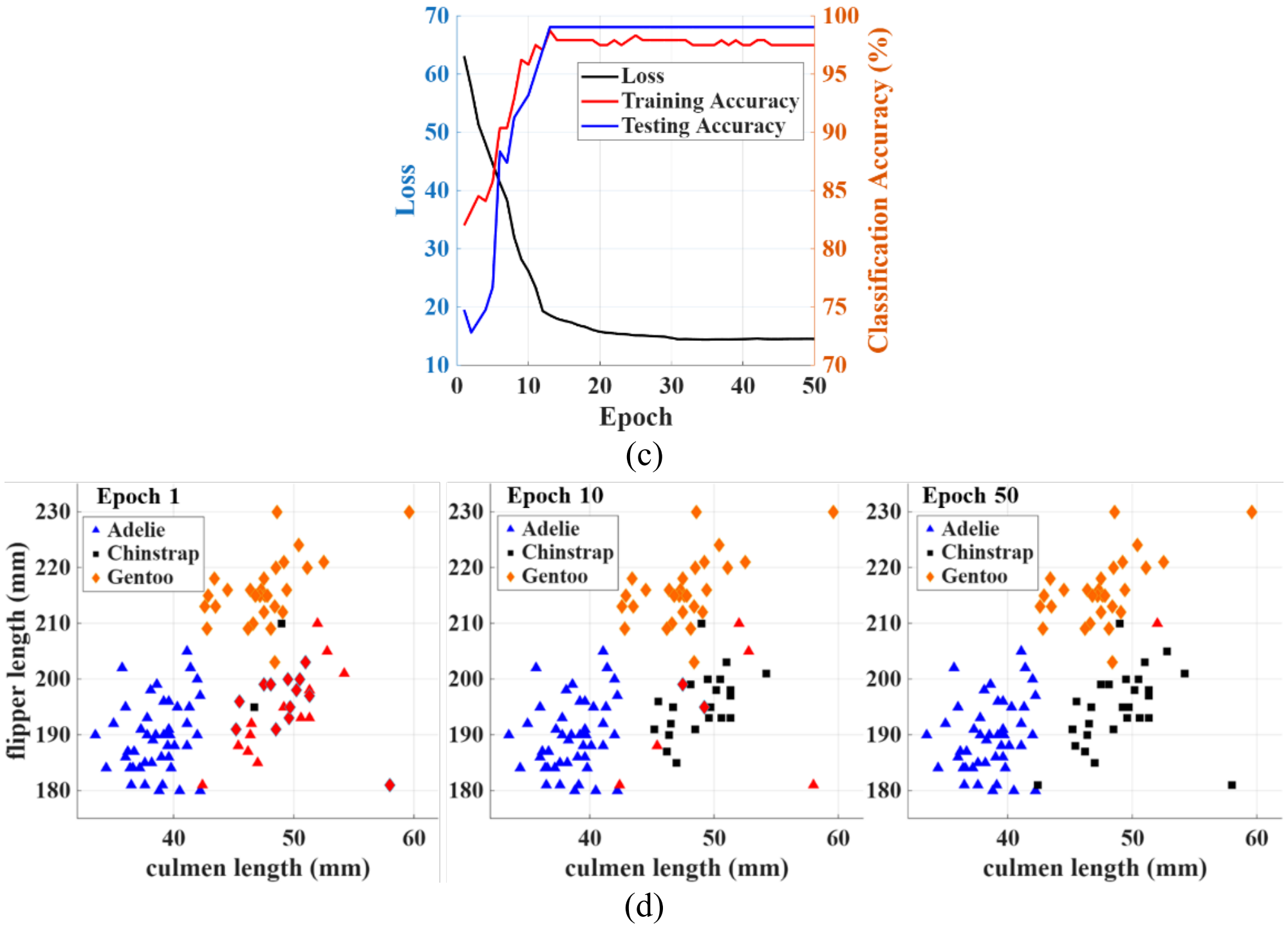

Fig. 4. The Wave-based Neuromorphic Circuit Network (WNC) as a penguin classifier. a) 2D projection of the Palmer Archipelago penguin data set. b) Trained WNC that classifies penguins: four physical features are input as voltage amplitudes and phases at the input plane ($V_1 = |l_{culmen}|\exp(jd_{culmen})$ and $V_2 = |l_{flipper}|\exp(jm_{bod})$), and each species is assigned to a node on the output plane. Predictions are decoded as the output node with the largest output power. c) Evolution of loss and accuracy on training and testing sets with epoch. d) Performance of the WNC on the testing dataset at epochs 1, 10, and 50. The incorrect classifications are filled in red.

Data classification is a stringent test used to benchmark an artificial neural network's performance. This section presents a WNC trained to classify the Palmer penguin dataset [33] through *in-situ* backpropagation. The learning task is to classify three penguin species based on four physical features: culmen length ($l_{culmen}$), culmen depth ($d_{culmen}$), flipper length ($l_{flipper}$), and body mass ($m_{bod}$). The 342 data points (distributed unevenly across the three species: Adelie-151, Chinstrap-68, and Gentoo-123) are randomly split into a training set (70%) and a testing set (30%). A two-dimensional projection of the four-dimensional data is shown in Fig. 4(a), illustrating the correlation between flipper and culmen length across the three species. Although the species are tightly grouped, they exhibit overlapping boundaries, with Adelie characterized by small flipper and culmen lengths and Gentoo by larger features.

The WNC is set up as follows. First, the input data is linearly normalized such that flipper length and culmen length lie in the range (0, 10] and are assigned to two voltage sources as amplitudes. The culmen depth and body mass are scaled to lie between [0° 180°] and applied to the two voltage sources as phases. Each source ($V_1 = |l_{culmen}|\exp(jd_{culmen})$ and $V_2 = |l_{flipper}|\exp(jm_{bod})$) is connected to a 40×20 grid along the input plane symmetrically (Fig. 4(b)) through a source resistance of 175 Ω, and separated from each other by 26 unit cells. Three nodes along the output plane are terminated in 175 Ω and each one is assigned to a species. The network's classification prediction is considered the species assigned to the node with the maximum output power. The remaining nodes along the periphery of the WNC are left open-circuited. As noted, input features are expressed as the amplitudes and phases of excitations, whereas the output predictions are measured in power quantities, resulting in a nonlinear mapping between inputs and outputs despite the medium's inherent linearity. The categorical cross-entropy loss function in Eq. (8) is used for training. The evolution of the loss and classification accuracy with each epoch is plotted in Fig. 4(c). The 2D distribution of phase delay across each unit cell of the optimized WNC is plotted in Fig. 4(b). The WNC achieves classification accuracy of 98% on the training set and 99% on the test set. For reference, a multinomial logistic regression classifier can achieve 100% test accuracy. The network's performance on the test set at epochs 1, 10, and 50 is shown in Fig. 4(d), with incorrectly classified labels marked in red. After just one epoch, the network can distinguish between Adelie and Gentoo but cannot identify Chinstrap. As the number of epochs increases, the network becomes more adept at discriminating among the three species.

### *4.3 Retraining a Damaged WNC*

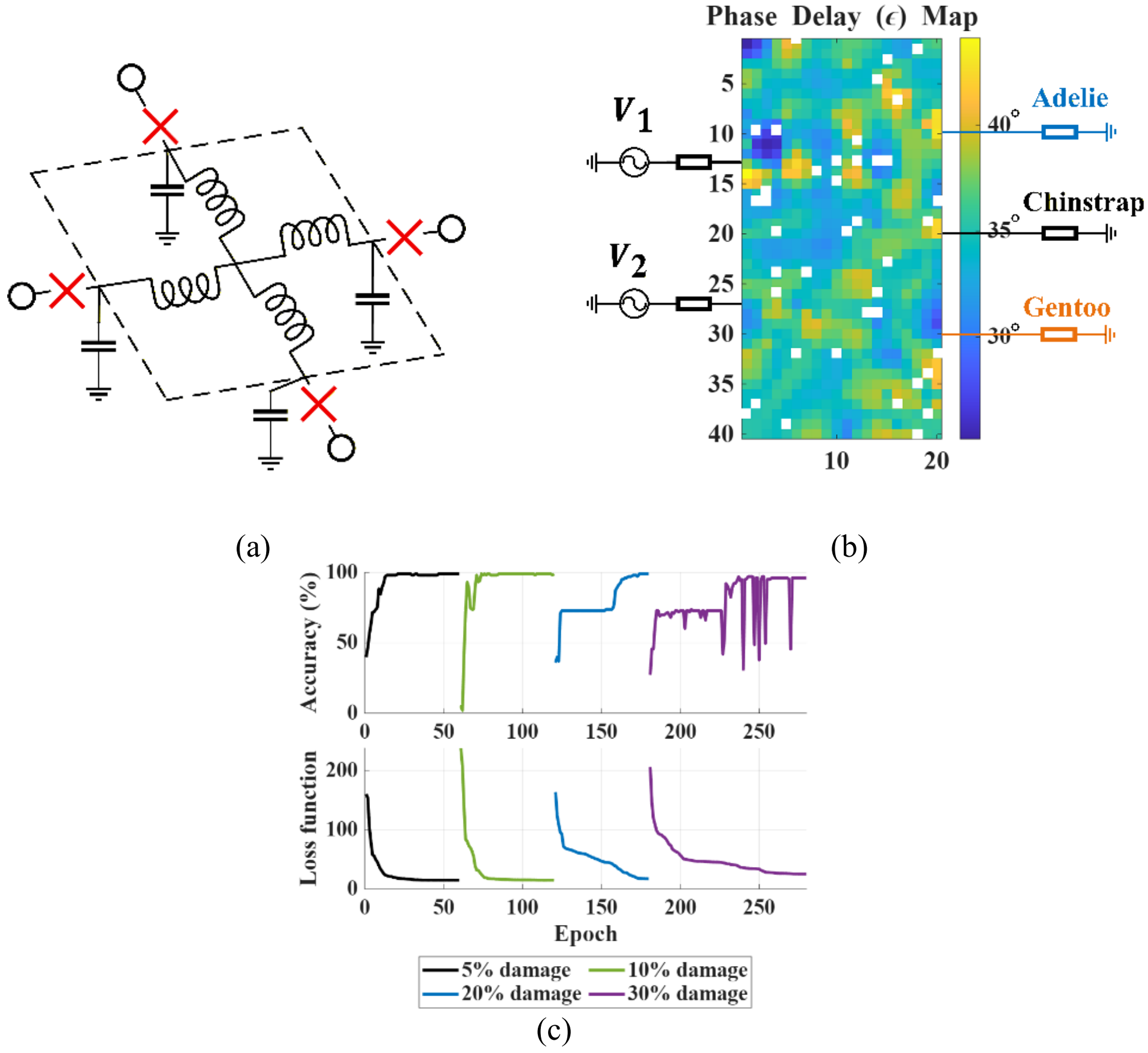

(c)

Fig. 5. Damaged Wave-based Neuromorphic Circuit Networks (WNCs) and their performance. a) Damaged TLIN unit cell (neuron) with its ports open-circuited. b) WNC with 5% of its unit cells subjected to damage (unit cells colored white are open-circuited), which is trained to classify the Palmer penguin dataset. b) Classification accuracy and loss function behavior with epoch for different levels of WNC damage.

Section 4.1 demonstrated that a WNC performing a three-output allosteric task, directing all the input power to a specified output node, can be easily retrained to direct power to another node. Another important feature of WNCs is their robustness to damage, due to co-located memory and processing, and the presence of multiple routing paths between input and output (distributed computation). Furthermore, *in-situ* backpropagation enables the network to be trained directly on the physical WNC hardware, thereby inherently accounting for defects and fabrication imperfections. This aligns with observations of biological systems, such as the brain, that can function and learn despite sustaining massive damage. In contrast, physical damage to a digital NN computing architecture typically results in operational failure. In other words, WNCs are likely to exhibit gradual degradation under localized damage compared to digital NNs [34,35].

To illustrate this robustness to damage, 5% of the unit cells in the trained WNC of section 4.2 are randomly selected and damaged. The WNC is then retrained to recover the original performance. In practice, lithographic errors, cracked transmission lines, or the failure of reactive components may render certain unit cells nonfunctional. Such defects are modeled here by electrically disconnecting the unit cells from the network. In other words, the four ports associated with each damaged cell are open-circuited, as depicted in Fig. 5(a). Fig. 5(b) shows an optimized WNC, with the damaged unit cells colored white, and Fig. 5(c) plots the loss and classification performance of the damaged WNC. The classification accuracy of the damaged WNC initially deteriorates to 45% before recovering to its original performance after retraining via in-situ backpropagation. The WNC is gradually subjected to further damage, and training in each case is depicted in Fig. 5(c). Surprisingly, the WNC can completely recover its performance even when a third of its area is damaged. It should be noted that the unit cells directly connected to the input and output nodes were not damaged.

### *4.4 Optical Character Recognition: Classification of Handwritten Digits*

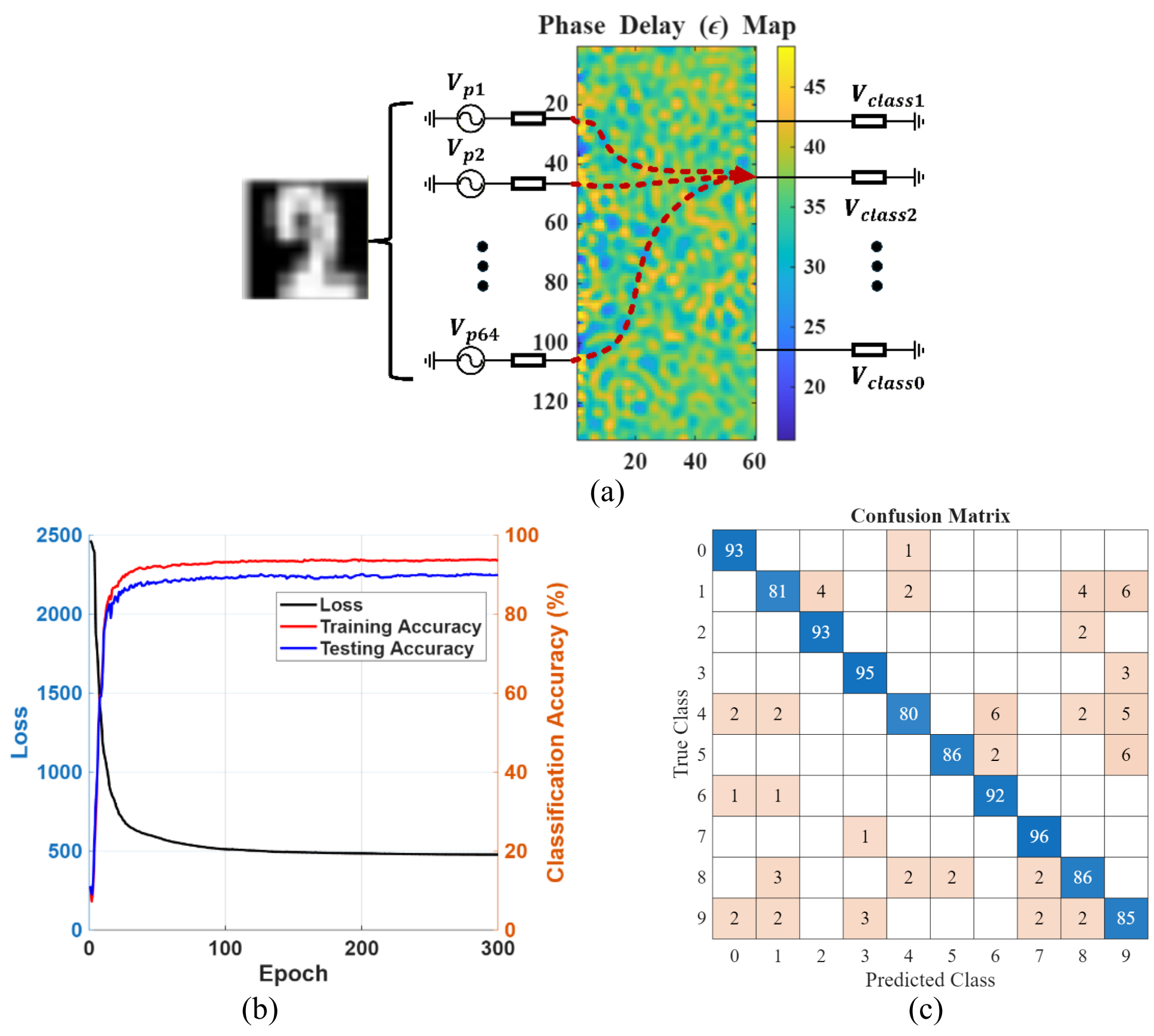

Fig. 6. Wave-based Neuromorphic Circuit Network (WNC) as a digit classifier. (a) WNC that classifies handwritten digits. The 2D image is input as an array of voltage source amplitudes, and the prediction is evaluated at the output plane. (b) Evolution of the loss function and classification accuracy with epoch. (c) Confusion matrix of the handwritten digit classification.

To demonstrate the effectiveness of WNCs on large multi-class datasets, a network is trained on a dataset of images (8×8 pixels) of handwritten digits [36], where each pixel's intensity can take an integer value in the range 0 to 16. The dataset was partitioned into 1914 images for training and 957 images for testing. Each pixel of an image is mapped to a corresponding voltage source at the input plane, and the pixel's raw intensity value is encoded as the magnitude of that source. Thus, the 64 pixels of each image are represented by 64 voltage sources, as illustrated in Fig. 6(a). The voltage sources are evenly distributed along the input plane of a 132×60 grid and connected in series with a source resistance of 175 Ω. Ten equidistant nodes along the output plane are assigned to each digit and terminated with a fixed impedance of 175 Ω. The remaining nodes on the periphery of the WNC are left open-circuited. As in Section 4.2, the network's prediction is taken to be the output node with the maximum output power. The categorical cross-entropy loss function is again used to train the WNC. The behavior of the loss function with epoch is plotted in Fig. 6(b). The 2D distribution of phase delay across each unit cell of the optimized WNC is plotted in Fig. 6(a), and the corresponding confusion matrix is shown in Fig. 6(c). Although the WNC achieves 91% classification accuracy over the test set, the confusion matrix indicates that it performs better on some digits than others. For reference, a logistic regression baseline on the same dataset achieved ~95% test accuracy. It is observed that not only the size of the network but also the manner of encoding and the spatial relationships between excited nodes and output nodes influence the WNC's performance, leaving room for improvement. The impact of input and output node placement on the functional capability of the WNC is left as future work [30].

## 5. Conclusion

This work introduces transmission-line (TLIN) metamaterials as a platform for neuromorphic computing. These wave-based neuromorphic circuit networks (WNCs) are physical neural networks that operate through steady-state wave-matter interactions. The proposed architecture consists of a 2D linear, reactive circuit network. It is composed of subwavelength TLIN-based unit cells, whose tunable shunt capacitances serve as programmable memory elements that store the learned input-output relationships. Inference is performed with high efficiency via wave propagation: scattering from tunable capacitances (susceptances), and interference across the 2D circuit network. The tunable capacitors discussed in this work could be implemented as switchable (discrete) capacitive states in non-volatile ferroelectric capacitors with non-destructive readout operation (NvCAPs) [37,38]. Unlike conventional ferroelectric capacitors, NvCAPs do not need to be reprogrammed after each readout. This saves energy and makes the read endurance of the device independent of its write endurance. The effect of the discrete tunability of shunt capacitances on the training and performance of the WNC is detailed in the Supplementary Section 2.

A key contribution of this work is the development of a scalable training methodology based on in-situ backpropagation. A physical (electric) realization of the backpropagation algorithm is derived using the adjoint variable method. It enables exact gradient computation using only two steady-state solutions of the physical network: the forward pass and adjoint pass (error backpropagation). Physically, it is a time reversal of the error signal through the same network used for the forward pass. It is shown how the gradient can be obtained from measurement of the local nodal voltages resulting from (a) the forward pass and (b) backpropagation of the error signal. *In-situ* gradient measurement makes WNCs well-suited for hardware implementation and scalable, since training is independent of the number of free parameters, and simulation-reality gaps associated with in-silico training can be avoided. Practical aspects, such as the adjoint excitation at the output ports and nodal-voltage measurements in hardware implementations, are non-trivial and warrant careful consideration. For laboratory experiments, near-field probing of planar microstrip LC grids has been demonstrated with both amplitude and phase fidelity [39,40]. An integrated CMOS implementation would address both challenges and enable scalability. The adjoint excitation can be synthesized using a shared on-chip reference oscillator with coherent multiport amplitude and phase control [41,42]. Nodal voltage measurements within each unit cell can be obtained via on-chip I/Q receivers co-located with each unit cell [43]. Both approaches (probes and on-chip I/Q receivers) are subject to measurement noise, which can degrade WNC performance. The effect of measurement noise on the training and performance of the WNC is detailed in Supplementary Section 3.

The capabilities of the proposed linear, wave-based neuromorphic circuits (WNCs) were demonstrated through representative tasks. These included allosteric power routing and multi-class classification using both structured datasets and image-based inputs. The results show that complex transformations between inputs and outputs can be realized using linear circuit networks. This is accomplished by encoding the information at the inputs and decoding it at the outputs via distinct physical quantities such as amplitude, phase, and power. In addition, the distributed and spatially recurrent nature of the network enables graceful degradation under localized damage, with performance recoverable through retraining using the same in-situ learning mechanism. Notably, the power-based output readout disregards the phase information of the output field. Substituting this approach with coherent I/Q detection would extend the output representation to the full complex plane for applications in which the output phase is informative.

These results highlight the potential of wave-based neuromorphic circuit networks as a new class of physical neural networks capable of performing computation. Such reconfigurable devices promise compact implementations, high energy efficiency, precision, and compatibility with existing CMOS fabrication processes or back-end-of-line (BEOL) processing. This enables scalable, adaptive implementations of high-speed, low-loss physical neural networks within existing CMOS-based electronic infrastructure for control and measurement systems. Future extensions of this framework include incorporating additional degrees of freedom within each unit through multiple tunable elements in tensor transmission line unit cell configurations [21] of shunt-node configuration (Fig. 7) and series-node configurations [44]. Defining a port across each variable element within the tensor unit cell preserves the local form of the gradient rule in Eq. (18). Nonlinear and time-varying components will also be considered, enabling increasingly sophisticated functionality, parallel processing through multi-frequency operation, and further reductions in device footprint. The adjoint gradient structure of (18) extends naturally to multi-frequency training, with each frequency contributing an independent pair of forward and adjoint solves. Furthermore, frequency multiplexing has been demonstrated as an effective strategy for increasing parallelism and throughput in neuromorphic photonic systems [45]. An analogous approach in the RF domain, enabled by time-varying elements within the WNC unit cell, is a direction we plan to pursue in future work. Finally, the efficient training using in-situ backpropagation reported in this work is also ideally suited for the design of multiport electromagnetic devices defined by complex-valued $K$ input-output pairs [19,20].

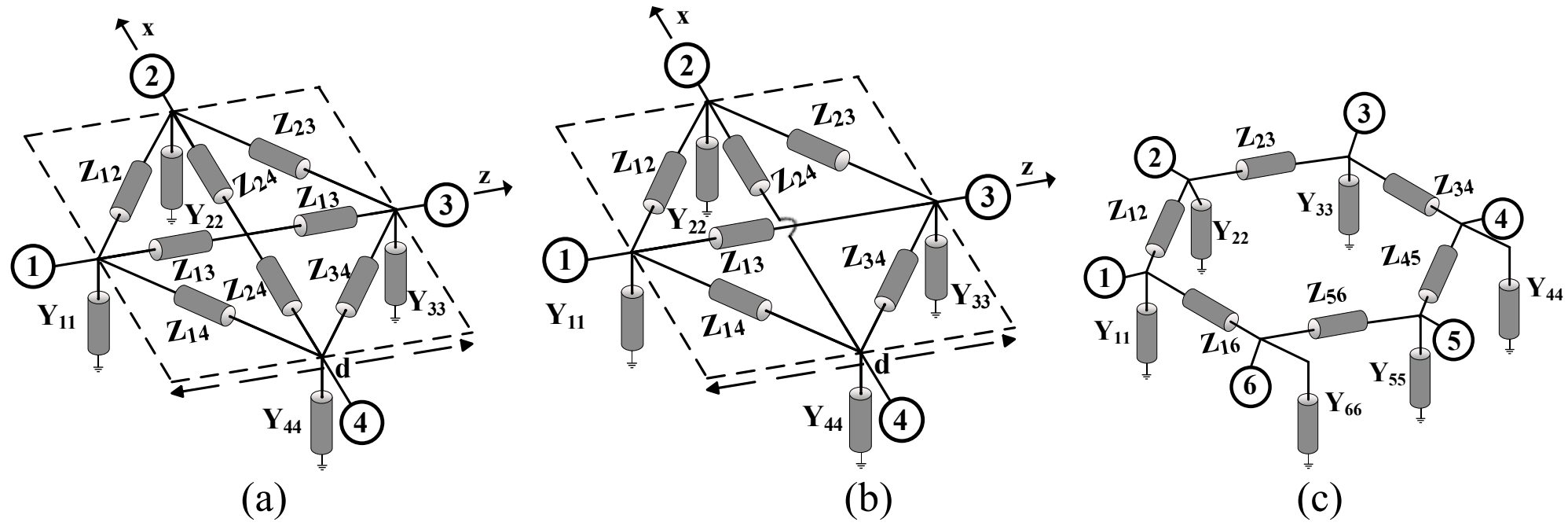

Fig. 7. Tensor transmission-line unit cell topologies with multiple tunable elements. a) Double bowtie unit cell. b) Fully-connected four-port unit cell c) Hexagonal unit cell

### Acknowledgements

This work was supported by the Air Force Office of Scientific Research (AFOSR) under Grant FA9550-24-1-0098.

See Supplement 1 for supporting content.

### References

1. Y. LeCun, Y. Bengio, and G. Hinton, "Deep learning," Nature **521**(7553), 436–444 (2015).
2. N. C. Thompson, K. Greenewald, K. Lee, and G. F. Manso, "The Computational Limits of Deep Learning," (2020).
3. V. Sze, Y.-H. Chen, T.-J. Yang, and J. S. Emer, "Efficient Processing of Deep Neural Networks: A Tutorial and Survey," Proc. IEEE **105**(12), 2295–2329 (2017).
4. A. N. Tait, M. A. Nahmias, B. J. Shastri, and P. R. Prucnal, "Broadcast and Weight: An Integrated Network For Scalable Photonic Spike Processing," J. Lightwave Technol. **32**(21), 4029–4041 (2014).
5. D. Rosenbluth, K. Kravtsov, M. P. Fok, and P. R. Prucnal, "A high performance photonic pulse processing device," Opt. Express **17**(25), 22767 (2009).
6. J. Feldmann, N. Youngblood, C. D. Wright, H. Bhaskaran, and W. H. P. Pernice, "All-optical spiking neurosynaptic networks with self-learning capabilities," Nature **569**(7755), 208–214 (2019).
7. O. Krestinskaya, A. P. James, and L. O. Chua, "Neuromemristive Circuits for Edge Computing: A Review," IEEE Trans. Neural Netw. Learning Syst. **31**(1), 4–23 (2020).
8. C. Li, D. Belkin, Y. Li, P. Yan, M. Hu, N. Ge, H. Jiang, E. Montgomery, P. Lin, Z. Wang, W. Song, J. P. Strachan, M. Barnell, Q. Wu, R. S. Williams, J. J. Yang, and Q. Xia, "Efficient and self-adaptive in-situ learning in multilayer memristor neural networks," Nat Commun **9**(1), 2385 (2018).
9. M. A. Zidan, J. P. Strachan, and W. D. Lu, "The future of electronics based on memristive systems," Nat Electron **1**(1), 22–29 (2018).
10. S. Pai, Z. Sun, T. W. Hughes, T. Park, B. Bartlett, I. A. D. Williamson, M. Minkov, M. Milanizadeh, N. Abebe, F. Morichetti, A. Melloni, S. Fan, O. Solgaard, and D. A. B. Miller, "Experimentally realized in situ backpropagation for deep learning in photonic neural networks," Science **380**(6643), 398–404 (2023).
11. S. Li and X. Mao, "Training all-mechanical neural networks for task learning through in situ backpropagation," Nat Commun **15**(1), 10528 (2024).
12. R. H. Lee, E. A. B. Mulder, and J. B. Hopkins, "Mechanical neural networks: Architected materials that learn behaviors," Sci. Robot. **7**(71), eabq7278 (2022).

13. R. Hamerly, L. Bernstein, A. Sludds, M. Soljačić, and D. Englund, "Large-Scale Optical Neural Networks Based on Photoelectric Multiplication," Phys. Rev. X **9**(2), 021032 (2019).
14. T. W. Hughes, M. Minkov, Y. Shi, and S. Fan, "Training of photonic neural networks through in situ backpropagation and gradient measurement," Optica **5**(7), 864 (2018).
15. S. Dillavou, M. Stern, A. J. Liu, and D. J. Durian, "Demonstration of Decentralized Physics-Driven Learning," Phys. Rev. Applied **18**(1), 014040 (2022).
16. X. Yan, J. H. Qian, V. K. Sangwan, and M. C. Hersam, "Progress and Challenges for Memtransistors in Neuromorphic Circuits and Systems," Advanced Materials **34**(48), 2108025 (2022).
17. A. Hirose, *Complex-Valued Neural Networks*, Studies in Computational Intelligence (Springer, 2012), **400**.
18. C. C. Wanjura and F. Marquardt, "Fully nonlinear neuromorphic computing with linear wave scattering," Nat. Phys. **20**(9), 1434–1440 (2024).
19. L. Szymanski, G. Gok, and A. Grbic, "Inverse Design of Multi-Input Multi-Output 2-D Metastructured Devices," IEEE Trans. Antennas Propagat. **70**(5), 3495–3505 (2022).
20. S. Thakkar, J. Ruiz-García, L. Szymanski, G. Gok, and A. Grbic, "Inverse Design of Perfectly Matched Metamaterials via Circuit-Based Surrogate Models and the Adjoint Method," IEEE Trans. Antennas Propagat. **73**(11), 9189–9202 (2025).
21. G. Gok and A. Grbic, "Tensor Transmission-Line Metamaterials," IEEE Trans. Antennas Propagat. **58**(5), 1559–1566 (2010).
22. M. G. Silveirinha, C. A. Fernandes, and J. R. Costa, "Electromagnetic Characterization of Textured Surfaces Formed by Metallic Pins," IEEE Trans. Antennas Propagat. **56**(2), 405–415 (2008).
23. J. Ruiz-García, S. Thakkar, G. Gok, and A. Grbic, "All-Metal Tensor Metamaterials: Characterization and Design," IEEE Transactions on Antennas and Propagation **72**(6), 5117–5128 (2024).
24. R. Sorrentino and F. Alimenti, "Waveguide Discontinuities," in *Encyclopedia of RF and Microwave Engineering*, K. Chang, ed., 1st ed. (Wiley, 2005).
25. M. S. S. Rahman, Y. Li, X. Yang, S. Chen, and A. Ozcan, "Massively parallel and universal approximation of nonlinear functions using diffractive processors," eLight **5**(1), 32 (2025).
26. C. Hammami, L. L. Magoarou, C. Monochristou, D. González-Ovejero, A. Momeni, R. Fleury, and P. del Hougne, "Expressivity of Programmable-Metasurface-Based Physical Neural Networks: Encoding Non-Linearity, Structural Non-Linearity, and Depth," (2026).
27. S. Thakkar and A. Grbic, "Fast Analysis of Metastructures using Multimodal Circuit-based Surrogate Models," in *2026 IEEE International Symposium on Antennas and Propagation and North American Radio Science Meeting (AP-S/CNC-USNC-URSI)* (IEEE, n.d.).
28. B. Scellier and Y. Bengio, "Equilibrium Propagation: Bridging the Gap between Energy-Based Models and Backpropagation," Front. Comput. Neurosci. **11**, 24 (2017).
29. M. Stern, D. Hexner, J. W. Rocks, and A. J. Liu, "Supervised Learning in Physical Networks: From Machine Learning to Learning Machines," Phys. Rev. X **11**(2), 021045 (2021).
30. J. W. Rocks, H. Ronellenfitsch, A. J. Liu, S. R. Nagel, and E. Katifori, "Limits of multifunctionality in tunable networks," Proc. Natl. Acad. Sci. U.S.A. **116**(7), 2506–2511 (2019).
31. J. W. Rocks, N. Pashine, I. Bischofberger, C. P. Goodrich, A. J. Liu, and S. R. Nagel, "Designing allostery-inspired response in mechanical networks," Proc. Natl. Acad. Sci. U.S.A. **114**(10), 2520–2525 (2017).

32. H. N. Motlagh, J. O. Wrabl, J. Li, and V. J. Hilser, "The ensemble nature of allostery," Nature **508**(7496), 331–339 (2014).
33. K. B. Gorman, T. D. Williams, and W. R. Fraser, "Ecological Sexual Dimorphism and Environmental Variability within a Community of Antarctic Penguins (Genus Pygoscelis)," PLoS ONE **9**(3), e90081 (2014).
34. A. Beygelzimer, G. Grinstein, R. Linsker, and I. Rish, "Improving network robustness by edge modification," Physica A: Statistical Mechanics and its Applications **357**(3–4), 593–612 (2005).
35. A. Kalampokis, C. Kotsavasiloglou, P. Argyrakis, and S. Baloyannis, "Robustness in biological neural networks," Physica A: Statistical Mechanics and its Applications **317**(3–4), 581–590 (2003).
36. C. K. E. Alpaydin, "Optical Recognition of Handwritten Digits," (1998).
37. S. Mukherjee, J. Bizindavyi, S. Clima, M. I. Popovici, X. Piao, K. Katcko, F. Catthoor, S. Yu, V. V. Afanas'ev, and J. Van Houdt, "Capacitive Memory Window With Non-Destructive Read in Ferroelectric Capacitors," IEEE Electron Device Lett. **44**(7), 1092–1095 (2023).
38. D. Yadav, S. Stathopoulos, P. Foster, A. Tsiamis, M. Awadein, H. Levene, and T. Prodromakis, "Multibit Ferroelectric HfZrO Memcapacitor for Non-Volatile Analogue Memory and Reconfigurable Electronics," Adv Funct Materials e31011 (2026).
39. Y. Li, Y. Sun, W. Zhu, Z. Guo, J. Jiang, T. Kariyado, H. Chen, and X. Hu, "Topological LC-circuits based on microstrips and observation of electromagnetic modes with orbital angular momentum," Nat Commun **9**(1), 4598 (2018).
40. C. Pfeiffer and A. Grbic, "A Printed, Broadband Luneburg Lens Antenna," IEEE Trans. Antennas Propagat. **58**(9), 3055–3059 (2010).
41. A. Natarajan, A. Komijani, and A. Hajimiri, "A fully integrated 24-GHz phased-array transmitter in CMOS," IEEE J. Solid-State Circuits **40**(12), 2502–2514 (2005).
42. A. Bhatta, J. Park, D. Baek, and J.-G. Kim, "A Multimode 28 GHz CMOS Fully Differential Beamforming IC for Phased Array Transceivers," Sensors **23**(13), 6124 (2023).
43. O. Inac, D. Shin, and G. M. Rebeiz, "A Phased Array RFIC with Built-In Self-Test Using an Integrated Vector Signal Analyzer," in *2011 IEEE Compound Semiconductor Integrated Circuit Symposium (CSICS)* (IEEE, 2011), pp. 1–3.
44. S. Akhtarzad and P. B. Johns, "Generalised elements for t.l.m. method of numerical analysis," Proc. Inst. Electr. Eng. UK **122**(12), 1349 (1975).
45. X. Xu, W. Han, M. Tan, Y. Sun, Y. Li, J. Wu, R. Morandotti, A. Mitchell, K. Xu, and D. J. Moss, "Neuromorphic computing using wavelength-division multiplexing," IEEE J. Select. Topics Quantum Electron. **29**(2: Optical Computing), 1–12 (2023).

## WAVE-BASED NEUROMORPHIC CIRCUIT NETWORKS: TUNABLE TRANSMISSION-LINE METAMATERIALS - SUPPLEMENTAL DOCUMENT

### 1. 2D Circuit Network Solver (CNS)

This section presents the construction of the large-scale admittance matrix that the 2D circuit network solver (CNS) uses to numerically simulate the tunable transmission-line metamaterial (the wave-based neuromorphic circuit network (WNC) considered in this work) during the forward pass (inference) and adjoint pass (error backpropagation). The CNS models the WNC (or, more generally, a 2D metamaterial that supports a single mode) as a large-scale admittance matrix that relates impressed Norton current sources (input data) to nodal voltages (prediction). A circuit network model of the WNC is constructed by interconnecting four-port admittance matrices, one for each TLIN unit cell, according to their physical arrangement, as shown in Fig. 1. The network is truncated by adding voltage sources (which can be zero or non-zero) connected in series with lumped impedances. These Thevenin sources are represented with Norton current sources and shunt admittances. In this way, the shunt admittances can be absorbed into the large-scale admittance matrix.

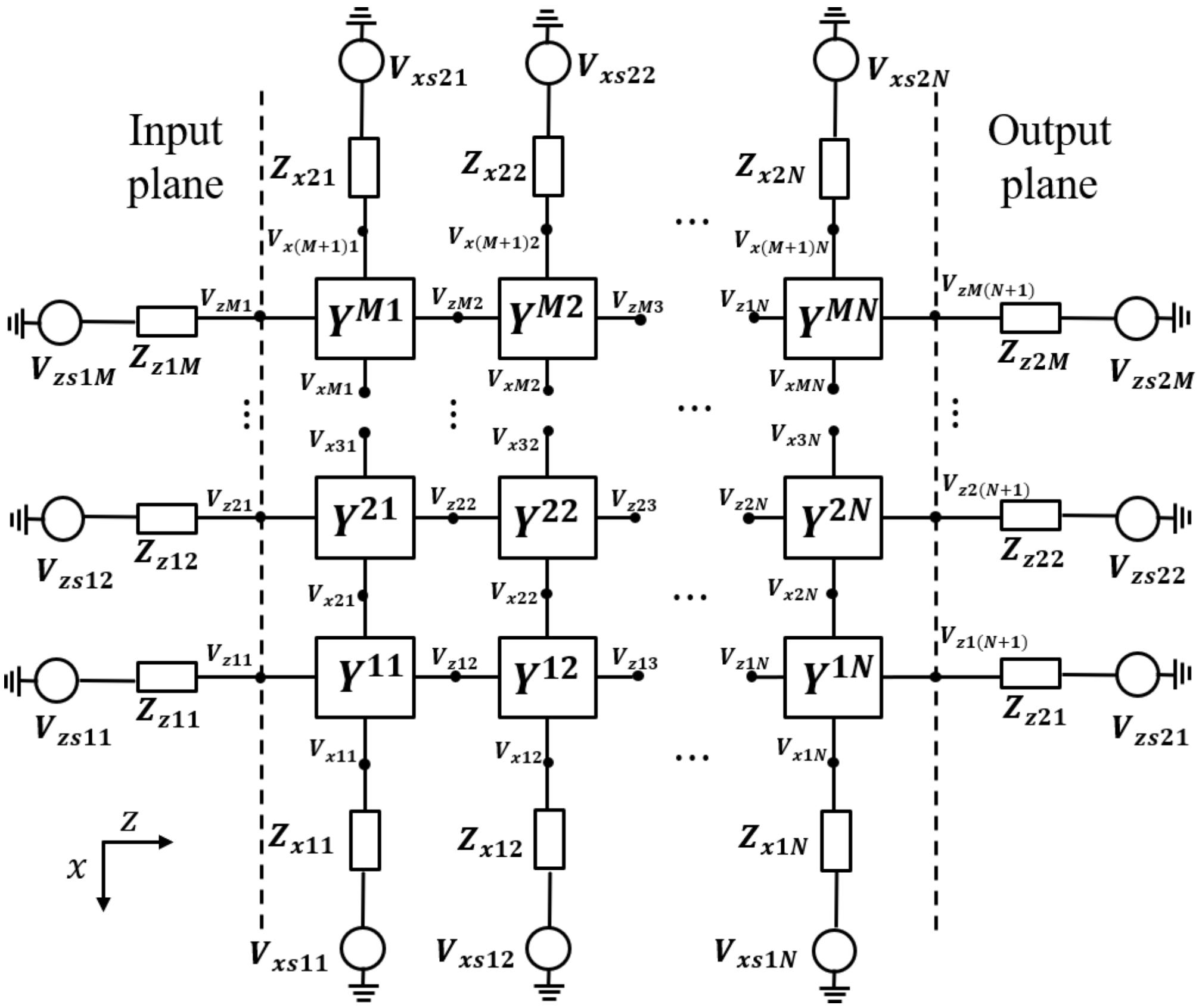

Fig. S1. The circuit network solver (CNS) models a metastructure consisting of an M×N grid of unit cells as a large-scale admittance matrix that relates impressed Norton current sources and nodal voltages. The large-scale admittance matrix is constructed by interconnecting four-port admittance matrices, one for each TLIN unit cell, and enforcing KCL at each node in the network. Boundary conditions are imposed using lumped impedances along the periphery [1].

Imposing Kirchoff's current law (KCL) at every interconnection between four-port admittance matrices (representing the unit cells) yields a sparse linear system of $n = 2\text{MN} + \text{M} + \text{N}$ equations whose solution is the voltage at every node in the network

$$\bar{\bar{Q}}\,\mathbf{v} = \mathbf{s}\,, \tag{S1}$$

where $\bar{\bar{Q}} \in \mathbb{C}^{n\times n}$ is the large-scale admittance (interaction) matrix that captures the internal structure of the WNC and the unit-cell interactions, $\mathbf{v}$ is a $\mathbb{C}^{n\times 1}$ vector containing all the nodal voltages (at the ports of the unit cells) in the network, and $\mathbf{s}$ is an $\mathbb{C}^{n\times 1}$ vector of Norton current sources impressed on the network. To determine the structure of $\bar{\bar{Q}}$, six types of nodes need to be examined: interior $V_x$ nodes, interior $V_z$ nodes and nodes on the four boundaries. The two types of interior nodes and a boundary node along the input plane are shown in Fig. S2. The three other boundary nodes can be examined in a manner similar to Fig. S2(c).

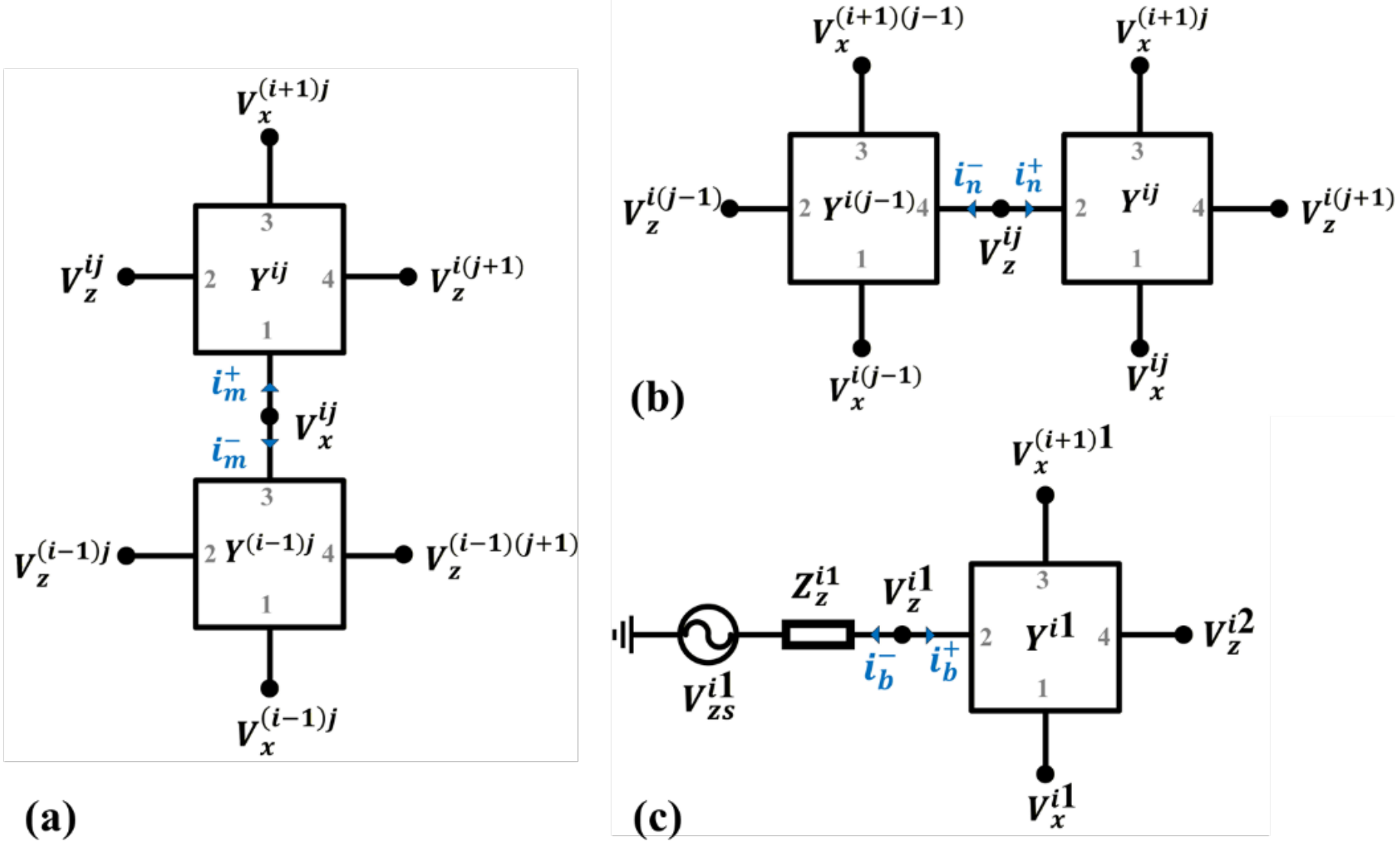

Fig. S2. Types of network nodes found in Fig. S1. a) Internal $V_x$ node. b) Internal $V_z$ node. c) Boundary node along the input plane. The remaining boundaries can be examined in a manner similar to (c).

Imposing KCL at a general interior $V_x^{ij}$ node shown in Fig. S2(a) yields,

$$\begin{aligned} V_x^{ij}\left(Y_{11}^{ij} + Y_{33}^{(i-1)j}\right) + V_x^{(i-1)j}Y_{31}^{(i-1)j} + V_x^{(i+1)j}Y_{13}^{ij} + V_z^{ij}Y_{12}^{ij} \\ + V_z^{(i-1)j}Y_{32}^{(i-1)j} + V_z^{(i-1)(j+1)}Y_{34}^{(i-1)j} + V_z^{i(j+1)}Y_{14}^{ij} = 0, \end{aligned} \tag{S2}$$

where $i$ and $j$ refer to the row and column address and the numeric subscripts refer to the unit cell ports (nodes). Similarly, applying KCL at a general interior $V_z^{ij}$ node shown in Fig. S2(b) yields,

$$\begin{aligned} V_z^{ij}\left(Y_{22}^{ij} + Y_{44}^{i(j-1)}\right) + V_z^{i(j-1)}Y_{42}^{i(j-1)} + V_z^{i(j+1)}Y_{24}^{ij} + V_x^{ij}Y_{21}^{ij} \\ + V_x^{(i+1)j}Y_{23}^{ij} + V_x^{(i+1)(j-1)}Y_{43}^{i(j-1)} + V_x^{i(j-1)}Y_{41}^{i(j-1)} = 0. \end{aligned} \tag{S3}$$

Applying KCL at the four types of boundary nodes leads to the following equations for the left boundary node $V_z^{i1}$ (Fig. S2(c))

$$V_z^{i1}\left(Y_{22}^{i1}+\frac{1}{Z_z^{i1}}\right)+V_z^{i2}Y_{24}^{i1}+V_x^{i1}Y_{21}^{i1}+V_x^{(i+1)1}Y_{23}^{i1}=\frac{V_{zs}^{i1}}{Z_z^{i1}}, \tag{S4}$$

the right boundary node $V_z^{i(N+1)}$

$$V_z^{i(N+1)}\left(Y_{44}^{iN}+\frac{1}{Z_z^{i2}}\right)+V_z^{iN}Y_{42}^{iN}+V_x^{iN}Y_{41}^{iN}+V_x^{(i+1)N}Y_{43}^{iN}=\frac{V_{zs}^{i2}}{Z_z^{i2}}, \tag{S5}$$

the bottom boundary node $V_x^{1j}$

$$V_x^{1j}\left(Y_{11}^{1j}+\frac{1}{Z_x^{1j}}\right)+V_x^{2j}Y_{13}^{1j}+V_z^{1j}Y_{12}^{1j}+V_z^{1(j+1)}Y_{14}^{1j}=\frac{V_{xs}^{1j}}{Z_x^{1j}}, \tag{S6}$$

And the top boundary node $V_x^{(M+1)j}$

$$V_x^{(M+1)j}\left(Y_{33}^{Mj}+\frac{1}{Z_x^{2j}}\right)+V_x^{Mj}Y_{31}^{Mj}+V_z^{Mj}Y_{32}^{Mj}+V_z^{M(j+1)}Y_{34}^{Mj}=\frac{V_{xs}^{2j}}{Z_x^{2j}}. \tag{S7}$$

This forms a linear system of $n$ equations, where $n$ is the number of nodes in the network. This system is written as a matrix equation in Eq (S1), where the admittance matrix entries are collected in the large-scale admittance matrix $\bar{\bar{Q}}$, and the Norton current sources on the RHS of Eq. (S2)-(S7) are gathered in $\bar{s}$. It should be noted that while $\bar{\bar{Q}}$ is an $n \times n$ matrix, there are a maximum of seven nonzero terms in each row, making it sparse and allowing rapid evaluation of the device response **v**. The four-port admittance matrix entries in Eq. (S2)-(S7) for each transmission-line-based unit cell shown in Fig. 1(a) are expressed as a function of the series inductances $L$ and shunt tunable capacitances $\mathrm{C}_{ij}$ in Eq. (4) and are rewritten here

$$Y_{11}=Y_{22}=Y_{33}=Y_{44}=\frac{1}{4}\left(Y_{\mathrm{C},ij}+\frac{3}{Z_a}\right), \tag{S8a}$$

$$Y_{21}=Y_{31}=Y_{41}=Y_{32}=Y_{42}=Y_{43}=-\frac{1}{4Z_a}, \tag{S8b}$$

where $Y_{\mathrm{C},ij}=j\omega\mathrm{C}_{ij}$, $Z_a=j\omega L$, $\omega$ is the radian frequency at 10 GHz, $\mathrm{C}_{ij}$ is the tunable shunt capacitance, and $L$ is set to 0.75 nH for all unit cells.

## 2. Effect of Discrete Tunability on WNCs

On-chip implementation of WNCs is important to scalability, realizing efficiency gains in inference, and enabling *in-situ* training. A major advantage of this architecture is compatibility with existing CMOS fabrication processes or back-end-of-line (BEOL) processing. This will enable the realization of tunable shunt capacitances using non-volatile ferroelectric capacitors (NvCAPs) [2,3]. They are characterized by their discrete tunability that can affect the training and performance of the WNC. To examine this, a WNC with shunt capacitances restricted to $2^b$ uniformly-spaced levels between 15-100 fF for b∈{1, 2, 3, 4} bits is trained on the Palmer Penguin dataset. Quantization-aware training was performed using a straight-through estimator: the continuous gradient from eq. (18) is used for the parameter update, and the capacitances are projected onto the nearest discrete level after each iteration. The evolution of the loss and (test) classification accuracy with epoch in each case is plotted in Fig. S3(a). With 3- and 4-bit capacitor resolution, classification performance is fully recovered relative to the continuous baseline (99.4%). Notably, 3-bit resolution is consistent with the number of stable states demonstrated in ferroelectric memcapacitors [3], indicating that current non-volatile capacitor technology is sufficient for the tasks demonstrated in this work. At 2 bits, a modest accuracy degradation to 82% is observed, and at b=1 bit, the performance further degrades to 78%. The 2D distribution of phase delay across each unit cell of the optimized WNC in the 1-bit case and 2-bit case is shown in Fig. S3(b)-(c). The network's performance on the dataset in the 1-bit case is shown in Fig. S3(d). The performance is consistent with a severely capacity-limited network that lacks the resolution to resolve the finer Adelie-Chinstrap boundary. This demonstration demonstrates the training routine's ability to accommodate practical effects.

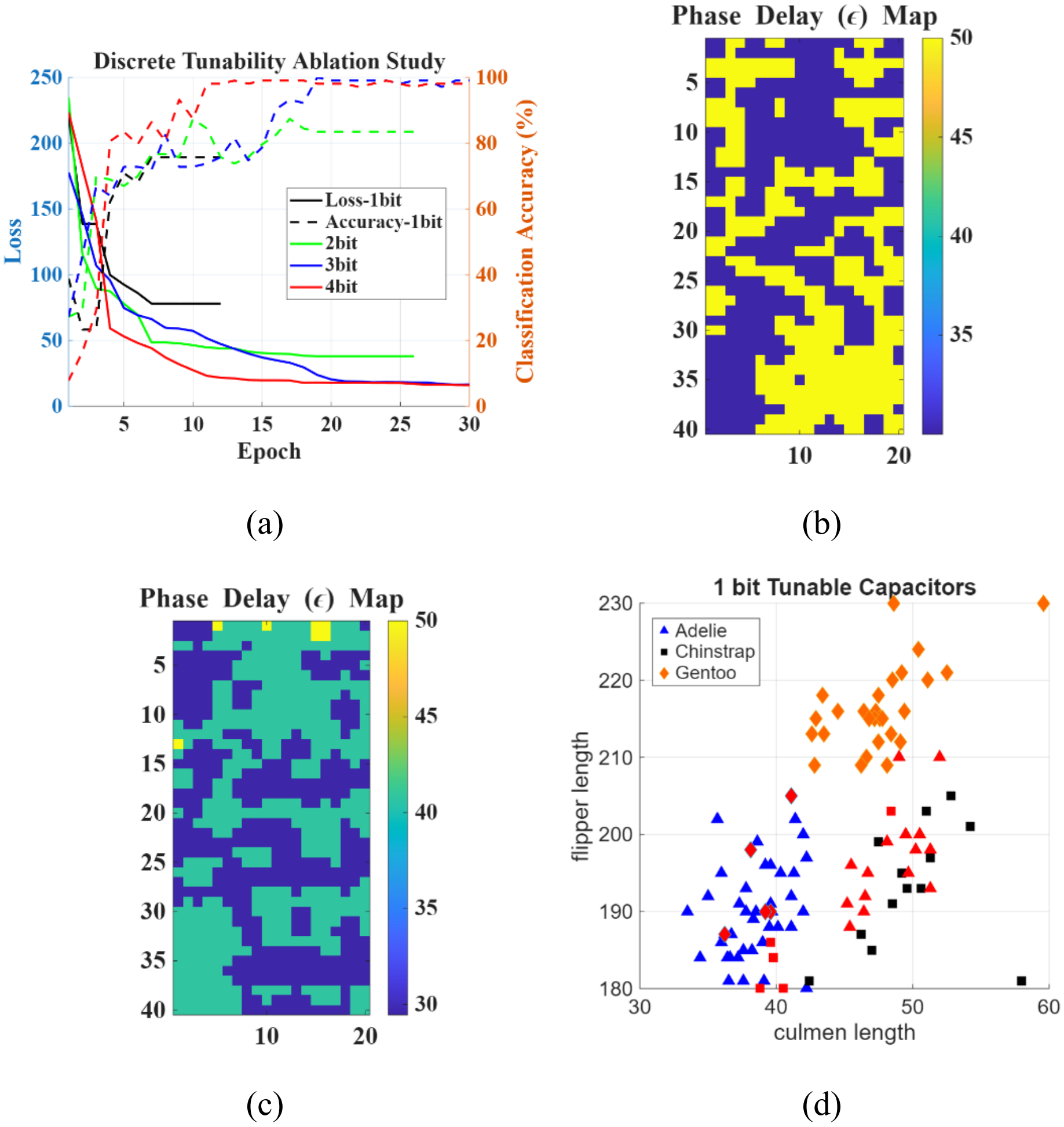

Fig. S3: Effect of discrete tunability in WNCs. a) Evolution of loss and accuracy on the testing set with epoch for b∈{1, 2, 3, 4} bits. 2D phase delay plot of an optimized WNC with b) 1-bit

tunable capacitors and c) 2-bit tunable capacitors. d) Performance of the optimized 1-bit WNC on the testing dataset. The incorrect classifications are filled in red.

## 3. Effect of Measurement Noise on WNCs

The measurement techniques mentioned in the main text, namely near-field probes and I/Q receivers for probing, are subject to measurement noise, which can degrade WNC performance. A simple additive Gaussian noise model is used to characterize this practical effect on WNC training. Additive Gaussian noise is added to each measured nodal voltage after both the forward and adjoint passes of the training algorithm. The noise power at each node is computed from the node's signal power and the predetermined SNR. Independent Gaussian noise samples are then added to the real and imaginary parts of each measured nodal voltage, and the gradient of Eq. (18) is evaluated from the noisy measurements. This models the finite SNR of the measurement hardware (typically 20-55 dB), and its effect on the accuracy of the gradient estimate for SNRs of 10, 20, 30, and 40 dB. The evolution of the loss and test classification accuracy with epoch in each case is plotted in Fig. S4(a). The classification accuracy is fully preserved at 20 dB SNR. At 10 dB SNR, accuracy degrades noticeably as the gradient estimate becomes corrupted. Near-field probing of microstrip grids and on-chip I/Q readout architectures can achieve above this threshold. The 2D phase delay plot at 10 dB SNR and the network's performance are plotted in Fig. S4(b)-(c).

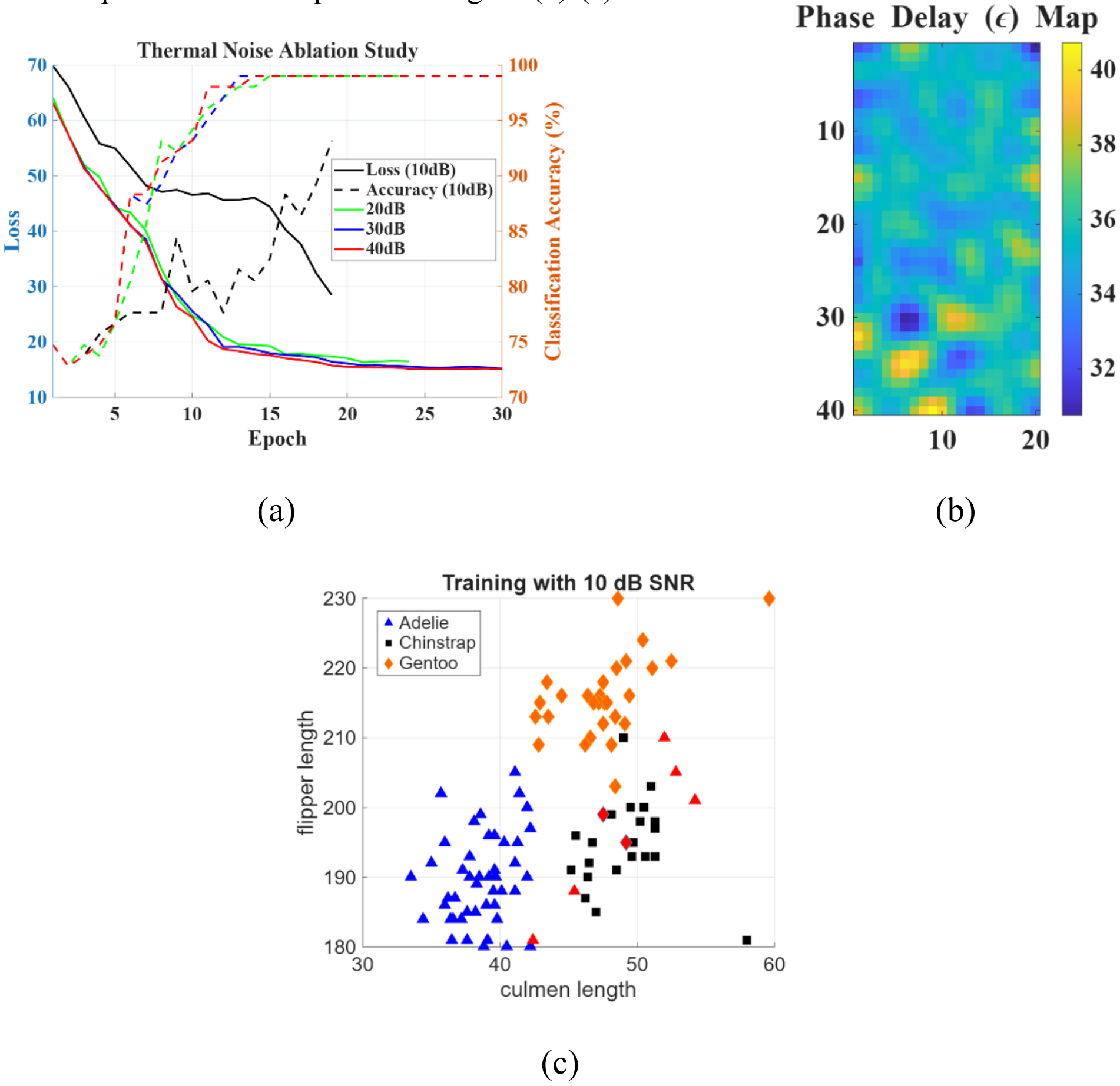

(a) (b)

(c)

Fig. S4. Effect of measurement noise on WNCs. a) Evolution of loss and accuracy on the testing set with epoch for SNR∈{10, 20, 30, 40} bits. b) 2D phase delay plot of an optimized WNC with 10 dB SNR on nodal voltage measurements. c) Performance of the optimized WNC on the testing dataset. The incorrect classifications are filled in red.

## 4. Effect of Loss on WNCs

We thank the reviewer for the suggestion. Loss is incorporated into the WNC's circuit network model by adding a resistance R in series with each unit cell's series inductance and a shunt conductance G in parallel with each tunable shunt capacitance, defining quality factors

$$Q_L = \frac{\omega L}{R} \qquad\qquad Q_C = \frac{\omega C}{G}. \tag{S9}$$

The dissipative elements account for losses experienced in a realized metastructure. The TLIN unit cell is depicted in Fig. S5. The unit cell consists of transmission-line interconnects on top of RO4003C substrate ($\boldsymbol{\varepsilon}$=3.3, height = 0.4 mm). The unit cell represents an effective distributed inductance of 1.4 nH, similar to the 1.5 nH value used in the paper. The HFSS eigenmode solver is used to extract a complex eigenfrequency of 10 GHz + $j$26 MHz for a phase delay of 35° across the unit cell. The imaginary part encompasses the loss mechanisms, and the real part is the resonant frequency, which remains unaffected by the addition of loss. As a result, the unit cell's quality factor is approximately 192. The quality factor of the shunt tunable capacitance is set to 30 based on estimated worst-case values from [3].

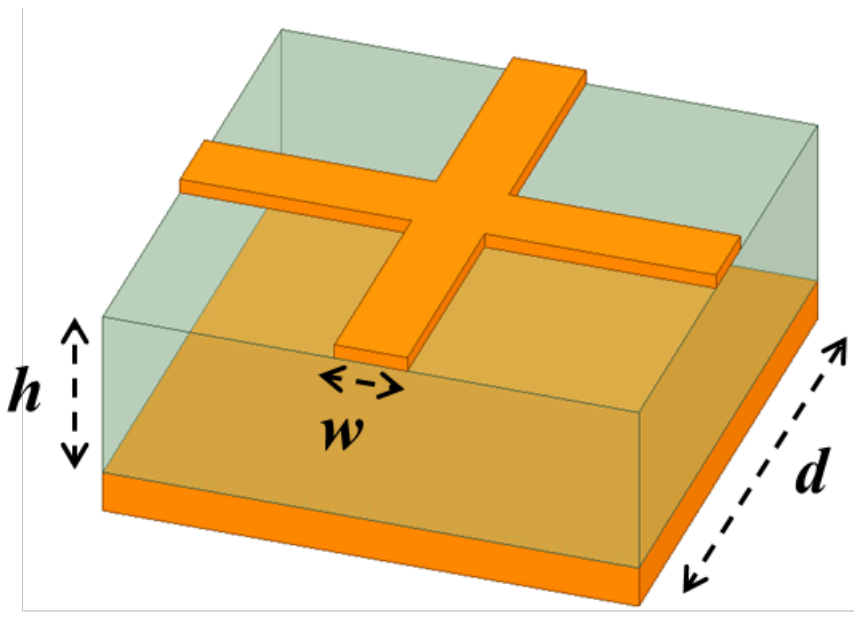

Fig. S5: Transmission-line unit cell having $d$ = 1.25 mm, $h$ = 0.4 mm, trace width $w$ = 0.17 mm. The transmission line interconnects provide an effective distributed inductance L = 0.7 nH and Q = 200.

The WNC is trained to classify penguins, as in section 4.2, with lossy components. The evolution of the loss and classification accuracy is shown in Fig. S6(a). It can be observed that the lossy WNC achieves a classification accuracy of 98%, which is just shy of the lossless case (see Fig. S6(b)-(c)). This confirms that graceful degradation can be expected in WNCs. Although the trained WNCs exhibit similar spatial features in Figs. S6(d)–(e), significant amplitude attenuation is evident in the lossy case.

We note that attenuation accumulates with propagation distance, and larger grids will therefore experience greater signal degradation. A promising approach to mitigate this effect is to incorporate time-varying elements within the unit cell, which can introduce parametric gain to compensate for propagation loss. This is an important direction in future work.

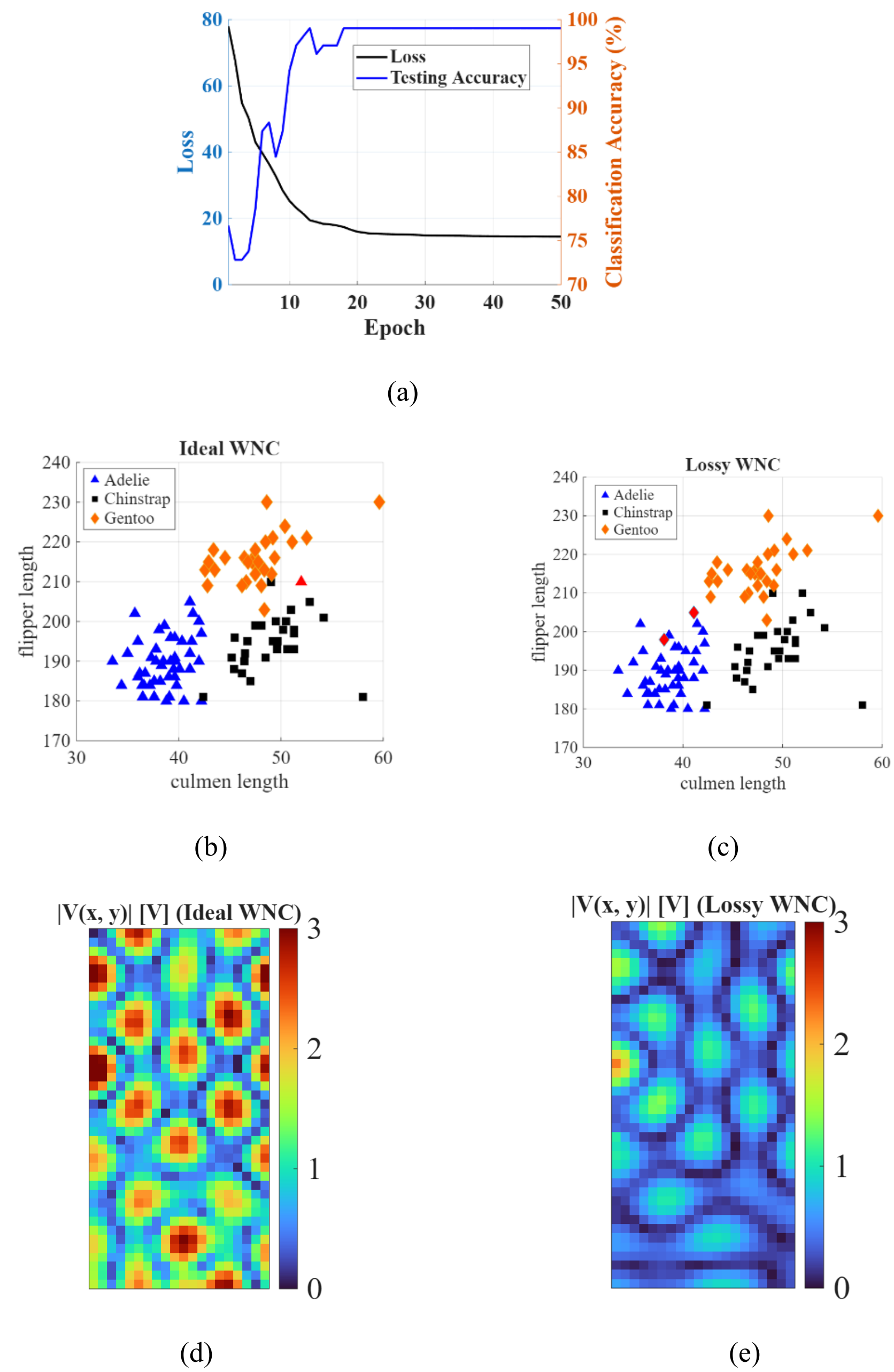

Fig. S6. Effect of loss on the performance of WNC that classified penguins (Section 4.3). Classification performance of a) Lossless WNC and b) Lossy WNC ($Q_C$ = 30, $Q_L$ =200). Simulation of the WNC for one case via the CNS for complex nodal voltages in the c) Lossless WNC and d) Lossy WNC.

## 5. Effect of Source Resolution on WNCs

In hardware implementation, the forward input sources and adjoint excitation sources are subject to finite DAC resolution, which quantizes their amplitudes and phases. This practical effect is incorporated by restricting each source amplitude and phase to $2^b$ discrete levels, where b refers to bits ∈{2, 3, 4}, prior to exciting the network in the forward and adjoint passes. The gradient is evaluated from the resulting nodal voltages. The bit depths and dynamic range considered here are motivated by [4], which demonstrates 5.6° phase resolution over 360°and uniform amplitude spacing over a 10x range (here, it is set to 1-10V to compare with section 4.2). The evolution of the loss and test classification accuracy with epoch in each case is plotted in Fig. S7(a). At 3 and 4 bits, classification accuracy reaches 97% and 98% respectively, compared to the 99% baseline shown in Section 4.2. The performance degrades appreciably for the 2-bit case to 78%. These results indicate that the DAC resolution requirement is comparable to the capacitance resolution demonstrated in Supplementary Section 2. The optimized WNC phase-delay maps for the 2-bit and 4-bit cases are plotted in Fig. S7(b)-(c). The classification performance in the 1-bit and 4-bit cases is shown in Fig. S7(d)-(e).

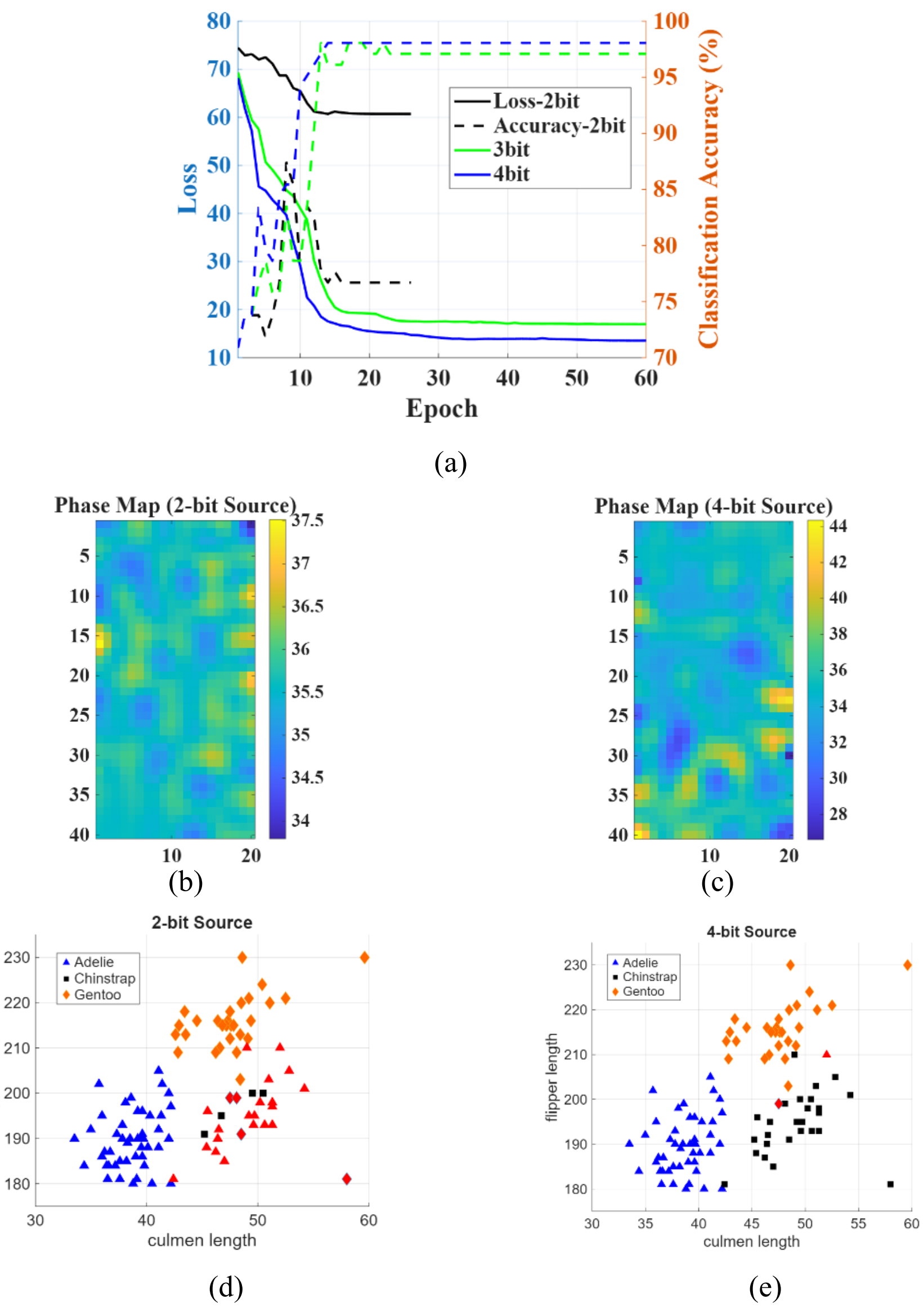

Fig. S7: Effect of source DAC resolution in WNCs. a) Evolution of loss and accuracy on the testing set with epoch for b∈{2, 3, 4} bits. 2D phase delay plot of an optimized WNC with b) 2-bit source DACs and c) 4-bit source DACs. Classification performance of the optimized d)

2-bit source DAC WNC and (e) 4-bit source DAC on the testing dataset. The incorrect classifications are filled in red.

## References

1. L. Szymanski, G. Gok, and A. Grbic, "Inverse Design of Multi-Input Multi-Output 2-D Metastructured Devices," IEEE Trans. Antennas Propagat. **70**, 3495–3505 (2022).
2. S. Mukherjee, J. Bizindavyi, S. Clima, M. I. Popovici, X. Piao, K. Katcko, F. Catthoor, S. Yu, V. V. Afanas'ev, and J. Van Houdt, "Capacitive Memory Window With Non-Destructive Read in Ferroelectric Capacitors," IEEE Electron Device Lett. **44**, 1092–1095 (2023).
3. D. Yadav, S. Stathopoulos, P. Foster, A. Tsiamis, M. Awadein, H. Levene, and T. Prodromakis, "Multibit Ferroelectric HfZrO Memcapacitor for Non-Volatile Analogue Memory and Reconfigurable Electronics," Adv Funct Materials e31011 (2026).
4. A. Bhatta, J. Park, D. Baek, and J.-G. Kim, "A Multimode 28 GHz CMOS Fully Differential Beamforming IC for Phased Array Transceivers," Sensors **23**, 6124 (2023).